\documentclass[11pt]{article}

\usepackage{jheppub}
\usepackage{tabularray}
\usepackage{diagbox}
\usepackage{hyperref}
\hypersetup{
	colorlinks,
	urlcolor=blue,
	linkcolor=magenta,
	citecolor=cyan,
	}
\usepackage{footmisc}
\usepackage{comment}

\newcommand{\sgn}{\text{sgn}}

\newcommand{\circw}{\mathbin{\tikz[baseline=-0.6ex]\draw[fill=white] (0,0) circle (0.5ex);}%
}

\usepackage{bm, bbm, bbold}
\usepackage{latexsym}
\usepackage{dcolumn}
\usepackage{amsmath,amsfonts,amssymb, mathrsfs, amsthm}
\usepackage[T1]{fontenc}

\usepackage{graphicx,epsfig}
\usepackage{environ}
\usepackage{mathtools}
\usepackage{braket}
\usepackage{fancyhdr}
\usepackage{hyperref}
\usepackage{graphicx,epstopdf}
\usepackage{tikz}
\usepackage{float}
\usepackage{framed}
\usepackage{soul}
\usetikzlibrary{positioning,decorations.markings, decorations.pathmorphing, calc}
\tikzset{snake it/.style={decorate, decoration=snake}}
\usetikzlibrary{angles,quotes}

\usepackage{stmaryrd}
\usepackage{slashed}
\usepackage{array,multirow}

\usepackage[framemethod=default]{mdframed}
\newmdenv[skipabove=7pt,
skipbelow=7pt,
rightline=false,
leftline=false,
topline=false,
bottomline=false,
backgroundcolor=gray!10,
linecolor=gray,
innerleftmargin=5pt,
innerrightmargin=5pt,
innertopmargin=5pt,
innerbottommargin=5pt,
leftmargin=0cm,
rightmargin=0cm,
linewidth=4pt]{eBox}

\def\e{\epsilon}

\def\p{\partial}

\newcommand{\be}{\begin{equation}}
\newcommand{\ee}{\end{equation}}
\newcommand{\bes}{\begin{equation*}}
\newcommand{\ees}{\end{equation*}}

\NewEnviron{derivation}{
\begin{framed}
\begin{center}
{\bf-: Derivation :-}\\
\end{center}
  \BODY

\end{framed}
}

\NewEnviron{eqn}{
\begin{align}
\begin{split}
  \BODY
\end{split}
\end{align}
}

\NewEnviron{eqn*}{
\begin{align*}
\begin{split}
  \BODY
\end{split}
\end{align*}
}

\newcommand{\nno}{\nonumber}
\newcommand{\intinf}{\int_{-\infty}^{\infty}} 
\newcommand{\intsinf}{\int_{0}^{\infty}} 

\usetikzlibrary{decorations.pathmorphing}	
\usetikzlibrary{decorations.markings}
\tikzset{
    sugra/.style={decorate, decoration={snake}, draw=black},
    scalarphi/.style={dashed,draw=black, postaction={decorate},
        },
    hwbou/.style={draw=blue, postaction={decorate}, ultra thick
        },
    vector/.style={draw=blue,decorate, decoration={snake}, draw},
	provector/.style={decorate, decoration={snake,amplitude=2.5pt}, draw},
	antivector/.style={decorate, decoration={snake,amplitude=-2.5pt}, draw},
   	 fermion/.style={draw=cyan, postaction={decorate},
        decoration={markings,mark=at position .55 with {\arrow[draw=black]{>}}}},
    fermionbar/.style={draw=cyan, postaction={decorate},
        decoration={markings,mark=at position .55 with {\arrow[draw=black]{<}}}},
    fermionnoarrow/.style={draw=black},
    gluon/.style={decorate, draw=red,
        decoration={coil,amplitude=4pt, segment length=5pt}},
    scalar/.style={dashed,draw=black, postaction={decorate},
        decoration={markings,mark=at position .55 with {\arrow[draw=black]{>}}}},
    scalarbar/.style={dashed,draw=black, postaction={decorate},
        decoration={markings,mark=at position .55 with {\arrow[draw=black]{<}}}},
    electron/.style={draw=black, postaction={decorate},
        decoration={markings,mark=at position .55 with {\arrow[draw=black]{>}}}},
    scalarnoarrow/.style={dashed, draw=black},
    electron/.style={draw=black, postaction={decorate},
        decoration={markings, mark=at position .55 with {\arrow[draw=black]{>}}}},
	bigvector/.style={decorate, decoration={snake, amplitude=4pt}, draw},
    photon/.style={draw=red, decorate, decoration={zigzag}, draw},
    higgs/.style={dashed, draw=black, postaction={decorate},
        },	
        goldstone/.style={draw=brown, postaction={decorate},
        },    
          ghost/.style={dashed, draw=magenta, postaction={decorate},
        decoration={markings, mark=at position .55 with {\arrow[draw=black]{>}}}
        },  
          antighost/.style={dashed, draw=magenta, postaction={decorate},
        decoration={markings, mark=at position .55 with {\arrow[draw=black]{<}}}
        },  
          mphoton/.style={decorate, decoration={snake}, draw=violet},
            realscalar/.style={draw=black}, 
           mgluon/.style={decorate, draw=blue,
        decoration={coil,amplitude=4pt, segment length=5pt}},
         weylfermion/.style={draw=orange, postaction={decorate},
        decoration={markings,mark=at position .55 with {\arrow[draw=black]{>}}}},
         weylfermionbar/.style={draw=orange, postaction={decorate},
        decoration={markings,mark=at position .55 with {\arrow[draw=black]{<}}}}, 
   	wboson/.style={draw=blue,decorate, decoration={snake,amplitude=4pt}, draw},  
    zboson/.style={draw=violet, decorate, decoration={snake}, draw},   
    lepton/.style={draw=black, postaction={decorate},
        decoration={markings,mark=at position .55 with {\arrow[draw=black]{>}}}},
    leptonbar/.style={draw=black, postaction={decorate},
        decoration={markings,mark=at position .55 with {\arrow[draw=black]{<}}}}, 
        graviton/.style={draw=blue,decorate, decoration={snake,amplitude=4pt}, draw},  
        gravitino/.style={draw=red, postaction={decorate},
        decoration={snake, markings, mark=at position .55 with {\arrow[draw=black]{>}}}},
    gravitinobar/.style={draw=red, postaction={decorate},
        decoration={snake, markings,mark=at position .55 with {\arrow[draw=black]{<}}} },    
}

\usepackage{cancel}
\usepackage{tikz, pgf}
\usetikzlibrary{shapes.misc}
\usetikzlibrary{shapes}
\usetikzlibrary{fit}
\usetikzlibrary{arrows}

\allowdisplaybreaks 

\begin{document}    
\title{On the simplicity of de Sitter correlators}
\author[b]{Chandramouli Chowdhury,}
\author[a,c,d]{Song He,}
\author[c,e]{Yong-Xiang Su}
\author[a,c,d]{and Dongyu Yang}
\affiliation[a]{School of Fundamental Physics and Mathematical Sciences, Hangzhou Institute for Advanced Study, UCAS, Hangzhou 310024, China}
\affiliation[b]{Mathematical Sciences and STAG Research Centre, University of Southampton,
Highfield, Southampton SO17 1BJ, United Kingdom}

\affiliation[c]{Institute of Theoretical Physics, Chinese Academy of Sciences, Beijing 100190, China}
\affiliation[d]{University of Chinese Academy of Sciences, Beijing 100049, China}
\affiliation[e]{School of Physical Sciences, University of Science and Technology of China, Hefei, Anhui 230026, China}
\emailAdd{C.Chowdhury@soton.ac.uk}
\emailAdd{songhe@itp.ac.cn}
\emailAdd{anonym@mail.ustc.edu.cn}
\emailAdd{yangdongyu24@mails.ucas.ac.cn}
\abstract{
Motivated by recent evidence that equal-time correlators can be simpler than the corresponding wavefunction coefficients, we study de Sitter correlators in conformally coupled $\phi^3$ theory directly. By inverting the momentum-space dressing rules, we derive a time integral representation for generic graphs and show that its natural building blocks are flat space correlators of fields and conjugate momenta. Among other things, this representation gives two useful recursive structures, one obtained by collapsing leaves and one by fusing lower-point graphs. In this representation several simplifications also become immediate. Graphs with an odd number of conjugate momentum insertions vanish, explaining the weight drop of odd-point correlators, melonic insertions collapse to lower complexity graphs and the leading behavior near total and partial-energy singularities is manifest, closely paralleling the flat space story. We then take a first step beyond the integrand and study integrated answers. For tree level families, in particular chains and stars, we find that the symbol alphabet of the correlator is smaller than that of the corresponding wavefunction, with the missing letters admitting a natural interpretation in terms of tubing data. These results support a correlator-first viewpoint for de Sitter observables: part of their simplicity appears to be intrinsic to the correlator itself, rather than inherited indirectly from the wavefunction.
}

\maketitle

\noindent
\flushbottom
\allowdisplaybreaks

\section{Introduction}
Cosmological correlators are the directly physical observables of the primordial universe \cite{Maldacena:2002vr,Arkani-Hamed:2015bza,Arkani-Hamed:2018kmz}.
They encode the late-time statistics of fluctuations generated during an approximate de Sitter phase \cite{Maldacena:2002vr,Baumann:2020dch}, and are the quantities ultimately extracted from the in-in formalism \cite{Schwinger:1960qe,Keldysh:1964ud,Weinberg:2005vy}.
The wavefunction of the universe is often an extremely useful auxiliary object, but it is the equal-time correlator that is directly tied to observables \cite{Ghosh:2014kba, Maldacena:2002vr,Arkani-Hamed:2018kmz,Baumann:2020dch}.
For this reason, understanding cosmological correlators on their own terms has become a central theme in recent years, both from the perspective of cosmology and from the broader modern program of uncovering hidden analytic structure in quantum field theory \cite{Raju:2012zr,Benincasa:2022gtd,Arkani-Hamed:2017fdk}.

A large part of the recent progress in this subject has come from the wavefunction side \cite{Hartle:1983ai,Arkani-Hamed:2017fdk}.
Wavefunction coefficients have proved to be particularly well suited for exposing singularities, flat space limits \cite{Raju:2012zr,Benincasa:2018ssx,Baumann:2021fxj}, bootstrap constraints \cite{Arkani-Hamed:2018kmz,Baumann:2019oyu,Baumann:2020dch} and positive-geometric structures, most notably through the cosmological polytope program and its extensions \cite{Arkani-Hamed:2017fdk,Hillman:2019wgh,Benincasa:2022gtd}.
This has led to the now familiar picture that cosmological observables, much like scattering amplitudes, are governed by structures far simpler than a direct sum over Feynman diagrams would suggest \cite{Arkani-Hamed:2017fdk,Hillman:2019wgh,Baumann:2021fxj}.
However, it has also raised a natural question: to what extent is this simplicity really a property of the wavefunction, and to what extent should it instead be attributed to the correlators themselves?

Recent work strongly suggests that the latter viewpoint deserves to be taken seriously.
In particular, it has become increasingly clear that correlators should not be regarded merely as quantities reconstructed from the wavefunction after the fact, but as observables with their own natural organization.
This has been sharpened from several directions, including direct and shadow-like formulations of in-in observables
\cite{Sleight:2020obc,DiPietro:2021sjt,Donath:2024utn,Chowdhury:2025nnk},
geometric constructions ranging from cosmological polytopes and cosmohedra to graph correlahedra and the correlatron
\cite{Arkani-Hamed:2017fdk,Benincasa:2024leu,Arkani-Hamed:2024jbp,Figueiredo:2025daa,Ardila-Mantilla:2026cbo,Glew:2026von},
and new relations between correlators and wavefunction coefficients
\cite{Stefanyszyn:2024wca}.
Perhaps most strikingly, a recent argument due to Arkani-Hamed \emph{et al.} shows that in flat space equal-time correlators are often simpler than wavefunctions precisely because they are obtained by integrating Feynman propagators over the full spacetime rather than over a half-space
\cite{Arkani-Hamed:2025mce}.
This observation gives a sharp conceptual reason why certain poles and redundancies present in wavefunction representations can disappear once one passes to the full correlator
\cite{Chowdhury:2023arc,Glew:2025arc,Chowdhury:2026upp}.

This correlator-first perspective is especially relevant for the present paper.
We study conformally coupled (CC) scalar theory with cubic interaction in de Sitter space, and our goal is to show that here too the correlator should be approached directly.
This is not merely a matter of presentation.
For the observables considered here, the de Sitter correlator is not obtained in any simple way from the flat space $\phi^3$ correlator by shifting external energies.
Instead, one is naturally led to a different time-domain representative, and it is in that representation that the simplifications of the de Sitter correlator become manifest.

Two recent developments make this perspective particularly concrete.
First, the in-out reformulation of in-in observables shows that cosmological correlators can, in suitable circumstances, be computed using a formalism much closer to ordinary Feynman propagators than might have been expected \cite{kamenev2023field,Donath:2024utn}.
Second, cosmological dressing rules provide a direct bridge between flat space graph data and de Sitter correlators, and in the CC $\phi^3$ setting they make it possible to identify the relevant pre-integrated building blocks explicitly \cite{Chowdhury:2025ohm}.
Together, these results suggest that de Sitter correlators admit direct representations in which recursion relations, singularity expansions, and other structural simplifications are already visible before the final integrations are performed.

Our aim in this paper is to develop this idea systematically for CC $\phi^3$ correlators.
We will show that they admit a natural direct representation, closely analogous in spirit to the emerging simplicity program for correlators but intrinsically adapted to de Sitter space.
This representation gives a transparent framework for discussing several kinds of structure at once: leaf recursions for trees, fusion recursions for more general topologies, melonic reductions, and expansions around the flat space limit and more general energy singularities.
It also provides a natural starting point for asking what survives after integration, in particular at the level of symbol alphabets.
One lesson that emerges is that the full correlator is often more economical than one might have expected from the corresponding wavefunction-like description, in line with the broader picture advocated in \cite{Chowdhury:2023arc,Glew:2025arc,Arkani-Hamed:2025mce}.

A further motivation comes from the symbolic structure of integrated cosmological observables.
For wavefunctions, symbols and alphabets have already revealed an unexpectedly rigid combinatorial organization \cite{Hillman:2019wgh,Baumann:2021fxj}.
For correlators, by contrast, this structure has been explored much less systematically.
If the correlator is indeed the more economical object, one should expect this economy to persist after integration: some letters natural from the wavefunction point of view should disappear, while the remaining ones should reorganize in a way better adapted to the full in-in observable.
The geometric picture of cosmological correlators \cite{Benincasa:2024leu} and the recent simplification results for correlators \cite{Chowdhury:2023arc,Arkani-Hamed:2025mce} strongly support this expectation, and one of our aims is to make it concrete in the CC $\phi^3$ setting.

The perspective we advocate is therefore modest but sharp.
We do not attempt here to solve CC $\phi^3$ correlators in complete generality after integration.
Instead, we identify the representation in which their structure is simplest to state, show that it organizes a non-trivial class of recursive, gluing, and singularity phenomena, and present first evidence that the same economy survives after integration.
In particular, the direct time-space representation naturally mixes field and conjugate momentum correlators, and this enlarged system closes under the operations relevant for the recursions studied below.

From this viewpoint, the integrated examples studied later in the paper should be regarded as probes of a broader conjectural picture.
If the correlator-first organization is the right one, then the integrated answers should not merely reproduce the wavefunction alphabet in disguise.
Rather, they should exhibit a smaller, or at least differently organized, set of letters with a direct graph-theoretic interpretation.
The examples we analyze provide first evidence for exactly this behavior.
Although our discussion of symbols is still exploratory, it points toward a version of the simplicity story in which the correlator comes with its own preferred alphabetic and geometric language.

\begin{figure}[H]
\hspace{-0.6cm}
\includegraphics[scale=.95]{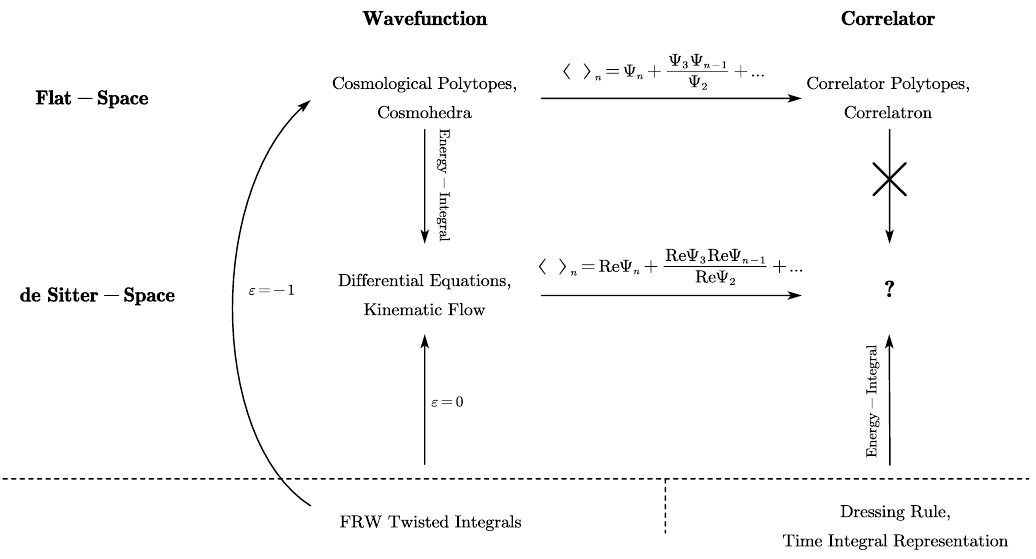}
\caption{Relations between wavefunctions and correlators in flat space and de Sitter space.}
\label{fig:RoadMap}
\end{figure}

Figure \ref{fig:RoadMap} summarizes the conceptual roadmap underlying this paper. The horizontal direction separates wavefunction-based and correlator-based organizations, while the vertical direction distinguishes flat space structures from their de Sitter counterparts.
On the wavefunction side, positive-geometric and symbol-based descriptions, such as cosmological polytopes and their extensions, have made singularities, flat space limits and alphabets especially transparent \cite{Arkani-Hamed:2017fdk,Benincasa:2018ssx,Hillman:2019wgh,Baumann:2021fxj}. The correlator side is more recent but suggests a parallel, and in some respects more economical, organization: correlator geometries and related flat space constructions indicate that equal-time correlators can possess their own intrinsic simplicity rather than merely inheriting it from the wavefunction \cite{Benincasa:2024leu,Chowdhury:2023arc,Glew:2025arc,Arkani-Hamed:2025mce}. The lower part of the figure indicates the route followed in the present work. Instead of obtaining the CC $\phi^3$ de Sitter correlator by a naive shift of a flat space correlator, we pass through the dressing rules and invert them into a direct time integral representation \cite{Donath:2024utn,Chowdhury:2025ohm}.
This places our construction in the lower-right corner of the diagram: a direct de Sitter correlator representation, designed to expose recursion relations and singularity expansions before integration, and to motivate the search for a smaller correlator alphabet after integration.
Related approaches based on twisted integrals, differential equations and kinematic flow provide complementary ways of organizing the same broader landscape \cite{Arkani-Hamed:2023decorr,Arkani-Hamed:2023kflow,Hang:2024xas,Baumann:2024loopflow,Baumann:2025qjx,Capuano:2025cde,Baumann:2026atn}.

The rest of the paper is organized as follows.
In Section~\ref{sec:directrep} we review the momentum-space dressing rules and derive from them a direct time integral representation for de Sitter correlators in CC $\phi^3$ theory.
In Section~\ref{sec:simplicity} we use this representation to expose several simplicity properties of the correlator, including vanishing theorems, leaf and fusion recursions, melonic reductions, and controlled expansions near energy singularities.
In Section~\ref{sec:symbol_simplicity} we turn to integrated answers and study their alphabets and symbols, emphasizing how they differ from the corresponding wavefunction data.
Taken together, these results support the view that de Sitter correlators are not merely secondary quantities reconstructed from the wavefunction, but observables with their own direct analytic structure and their own version of the cosmological simplicity story \cite{Lee:2023kno,Benincasa:2024leu,Donath:2024utn,Chowdhury:2025ohm,Arkani-Hamed:2025mce}.

\section{Direct Representations of de Sitter Correlators}\label{sec:directrep}

In this section we review the momentum-space dressing rules for CC $\phi^3$ correlators and derive the time integral representation that will be used throughout the rest of the paper.
Our main goal is to make explicit the variables in which the correlator admits a direct description, without first passing through the wavefunction.

\subsection{Dressing Rules}\label{recap:momspacedress}
In this subsection we briefly review the dressing rules of \cite{Chowdhury:2025ohm}, which express de Sitter correlators as integrals over flat space Feynman diagrams dressed by auxiliary kernels.
These kernels depend on the theory and on the interaction under consideration, and the resulting representation provides an efficient momentum-space starting point for the rest of our analysis.
We summarize the general prescription and then illustrate it with the simplest exchange graph.

Consider a Feynman diagram corresponding to a single exchange graph without energy conservation. To obtain the dS correlator we add two additional (auxiliary) propagators  at each vertex each carrying energy $p_1, p_2$. Each of these propagators depend on the external energy entering the vertex, which are denoted by $x_1, x_2$ respectively. Finally, along with the usual spatial momentum conservation, there is energy conservation between the energy of the internal legs and the auxiliary propagators at every vertex. Graphically this is denoted as,
\begin{eqn}\label{eq:dressingrev1}
\begin{tikzpicture}[baseline]
\draw (-1.5, 1) -- (-1, 0);
\draw (-1.5, -1) -- (-1, 0);
\draw[scalar,solid] (-1, 0) -- (1,0);
\draw (1.5, 1) -- (1, 0);
\draw (1.5, -1) -- (1, 0);
\draw[scalar, black] (-1, 0) -- (0, -1);
\draw[scalar, black] (1, 0) -- (0, -1);
\node at (0, 0.25) {$p$};
\node at (-0.7, -0.7) {$p_1$};
\node at (0.7, -0.7) {$p_2$};
\node at (-1.25, 0) {$x_1$};
\node at (1.25, 0) {$x_2$};
\end{tikzpicture}
= \intinf \mathrm{d}p_1 \mathrm{d}p_2 \mathrm{d}p K(x_1, p_1) K(x_2, p_2) \frac{\delta(p_1 + p) \delta(p_2 - p) \delta(p_1 + p_2)}{p^2 + k^2}
\end{eqn}
where $\vec k$ is the spatial momentum carried by the internal leg. We have suppressed the spatial momentum conserving delta functions. The kernel $K(x, p)$ depends on the theory \cite{Chowdhury:2025ohm} and we summarize the ones we shall be discussing in this paper below. This procedure can be generalized to all graphs at any perturbative order. For Feynman diagrams at loop level further caution must be exercised since the integrals can be divergent \cite{Senatore:2009cf, Chowdhury:2023arc} and we postpone that discussion to future work.

For the theories of interest in this paper the kernels are given as follows.
\begin{enumerate}
\item \underline{CC $\phi^4$ theory}: This is equivalent to massless theories in flat space with time independent interactions. There is a single kind of Kernel given as
\begin{eqn}\label{eq:flatkernel}
K(x, p) = \frac{-2x}{p^2 + x^2}. 
\end{eqn}

\item \underline{CC $\phi^3$ theory}:  This is equivalent to massless theories in flat space with time dependent interactions \cite{Arkani-Hamed:2017fdk}. For these theories there are two different kernels which are denoted by {\it dashed} and {\it dotted} propagators. 

The Kernel corresponding to dashed propagators is given as, 
\begin{eqn}\label{eq:dashprop-review}
K_{\rm dash}(x, p) = -2\mathrm{i} \intsinf \mathrm{d}s \frac{p}{p^2 + (s + x)^2}
\end{eqn}
hence these are generically logarithmic in the energies. For this theory, the kernels must satisfy an additional constraint requiring that the number of dashed propagators be even.

The Kernel corresponding to the dotted propagator is much simpler,
\begin{eqn}\label{eq:dotprop-review}
K_{\rm dot}(x, p) = -\pi
\end{eqn}
While a constant propagator might initially seem surprising, it arises naturally from the expression for the correlator in terms of wavefunction coefficients; in particular, it corresponds to the real part of the 3-point wavefunction coefficient in conformally coupled $\phi^3$ theory.
\end{enumerate}

We provide an example for the 2-site graph which illustrates the utility of this representation. The dS correlator for the $\phi^3$ theory is given via the sum of 2 graphs
\begin{align}
\begin{tikzpicture}[baseline]
\draw (-1.5, 1) -- (-1, 0);
\draw (-1.5, -1) -- (-1, 0);
\draw[scalar,solid] (-1, 0) -- (1,0);
\draw (1.5, 1) -- (1, 0);
\draw (1.5, -1) -- (1, 0);
\draw[scalar, black] (-1, 0) -- (0, -1);
\draw[scalar, black] (1, 0) -- (0, -1);
\node at (0, 0.25) {$p$};
\node at (-0.7, -0.7) {$p_1$};
\node at (0.7, -0.7) {$p_2$};
\node at (-1.25, 0) {$x_1$};
\node at (1.25, 0) {$x_2$};
\end{tikzpicture} &= (-4\lambda^2) \intsinf \mathrm{d}s_1 \mathrm{d}s_2 \intinf \mathrm{d}p \mathrm{d}p_1 \mathrm{d}p_2 \frac{p_1 p_2 \delta(p_1 + p) \delta(p_2 - p) }{\big(p_1^2 + (s_1 + x_1)^2 \big)\big(p_2^2 + (s_2 + x_2)^2 \big)(p^2 + k^2) }, \nno\\
\begin{tikzpicture}[baseline]
\draw (-1.5, 1) -- (-1, 0);
\draw (-1.5, -1) -- (-1, 0);
\draw[scalar,solid] (-1, 0) -- (1,0);
\draw (1.5, 1) -- (1, 0);
\draw (1.5, -1) -- (1, 0);
\draw[black, dotted] (-1, 0) -- (0, -1);
\draw[black, dotted] (1, 0) -- (0, -1);
\node at (0, 0.25) {$p$};
\node at (-1.25, 0) {$x_1$};
\node at (1.25, 0) {$x_2$};
\end{tikzpicture} &= \pi^2 \lambda^2 \intinf \mathrm{d}p \frac{1}{p^2 + k^2}
\end{align}
Their sum reproduces the standard tree level four-point correlator in conformally coupled \(\phi^3\) theory \cite{Arkani-Hamed:2015bza}.

The 2-site example is the prototype for the momentum-space representations used later in the paper.
More generally, every correlator in conformally coupled \(\phi^3\) theory is obtained by summing over all allowed dashed/dotted dressings of the corresponding flat space diagram, subject to the condition that the number of dashed auxiliary propagators is even, and then integrating over the remaining internal and auxiliary energies.
In the next subsection we will invert this representation and recast it as a direct time integral, which is the form in which the simplicity properties of the correlator become most transparent.

\subsection{Time Integral Representation}\label{ssec:timeint}
It was shown in \cite{Chowdhury:2025nnk} how momentum-space dressing rules in flat space can be obtained from a corresponding time integral representation via the in-out formalism \cite{Donath:2024utn}.
Here we proceed in the opposite direction and derive a time integral representation for CC $\phi^3$ correlators in de Sitter space by inverting the momentum-space formulae of \cite{Chowdhury:2025ohm}.


We first review the time integral representation for flat space correlators with an example. The formula for a $W$-pt correlator of field insertions in flat space is given as
\begin{eqn}
\braket{G}_W = \intinf \prod_{i = 1}^W \mathrm{d}t_i e^{- x_i |t_i|} \prod_{e \in E} G_F(t, t'; y_e)
\end{eqn}
where $E$ denotes the edges of the graph and $\{t, t'\} \in \{t_1, \cdots, t_W\}$. This is the standard expression for the time-ordered correlators in perturbation theory, written as a product of Feynman propagators. For loop level graphs one must additionally integrate over the $y$-variables (see appendix \ref{app:loop} for an example). To denoting individual graphs, we introduce a shorthand notation for the vertices $\circw$. 

As an example consider the 2-site graph,
\begin{eqn}\label{2sitefltatime1}
\begin{tikzpicture}[baseline]
\draw (-0.5, 0) -- (0.5,0);
\node at (-0.5, 0) {$\circw$};
\node at (0.5, 0) {$\circw$};
\node at (-0.5, -0.25) {$x_1$};
\node at (0.5, -0.25) {$x_2$}; 
\node at (0, +0.25) {$y$};
\end{tikzpicture}
= \intinf \mathrm{d}t_1 \mathrm{d}t_2 e^{- x_1 |t_1|} e^{- x_2 |t_2|} G_F(t_1, t_2;  y)
\end{eqn}
where $G_F(t_1, t_2; y) = \frac{e^{- y |t_1 - t_2|}}{y} $ is the Feynman propagator for a massless field. To derive the momentum-space dressing rules of \cite{Chowdhury:2025ohm} we use the Fourier space representation for these propagators, 
\begin{eqn}\label{eq:feynprop1}
e^{- x|t|} &=  \intinf \frac{\mathrm{d}p}{\pi} \frac{x}{p^2 + x^2} e^{\mathrm{i} p t}, \qquad 
G_F(t_1, t_2; y) = \intinf \frac{\mathrm{d}q}{\pi} \frac{1}{q^2 + y^2} e^{\mathrm{i} q (t_1 - t_2)}
\end{eqn}
Plugging these into \eqref{2sitefltatime1} and performing the time integrals we get,
\begin{eqn}\label{2siteflatmomspace}
\begin{tikzpicture}[baseline]
\draw (-0.5, 0) -- (0.5,0);
\node at (-0.5, 0) {$\circw$};
\node at (0.5, 0) {$\circw$};
\node at (-0.5, -0.25) {$x_1$};
\node at (0.5, -0.25) {$x_2$}; 
\node at (0, +0.25) {$y$};
\end{tikzpicture}
&=\frac{1}{\pi^3} \intinf \mathrm{d}p_1 \mathrm{d}p_2 \mathrm{d}q \frac{ x_1 x_2}{(p_1^2 + x_1^2)(p_2^2 + x_2^2)}  \intinf \mathrm{d}t_1 \mathrm{d}t_2 e^{\mathrm{i} (p_1 + q) t_1} e^{\mathrm{i} (p_2 - q) t_2} \\
&= \frac{4}{\pi} \intinf \mathrm{d}p_1 \mathrm{d}p_2 \mathrm{d}q \frac{ x_1 x_2}{(p_1^2 + x_1^2)(p_2^2 + x_2^2)} \frac{\delta(p_1 + q) \delta(p_2 - q)}{q^2 + y^2}
\end{eqn}
This resembles the general form of the answer given in \eqref{eq:dressingrev1} for the specific choice of Kernel given in \eqref{eq:flatkernel}. The variables $p_1, p_2, q$ are to be identified with the ``auxiliary energies'' in the language of \cite{Chowdhury:2025ohm} and the $\delta$ functions indicate the conservation of auxiliary energies at each vertex (analogous to energy conservation for scattering amplitudes). It should be obvious from this example that this procedure can be generalized to any graph.

For theories in dS, specifically the CC scalar fields with $\phi^3$ interaction, we can reverse engineer the dressing rule to obtain the time integral representation. As reviewed in the previous section  \ref{recap:momspacedress} there are now two kinds of auxiliary propagators: dashed and dotted. We first provide a time integral representation for the dashed propagator by retracing the steps for the 2-site graph. Using the dressing rule we have the following representation in momentum-space (we set the coupling constants to $1$),
\begin{eqn}
\begin{tikzpicture}[baseline]
\draw (-0.75, 0.5) -- (-0.5, 0);
\draw (-0.75, -0.5) -- (-0.5, 0);
\draw (-0.5, 0) -- (0.5, 0);
\draw (0.75, 0.5) -- (0.5, 0);
\draw (0.75, -0.5) -- (0.5, 0);
\draw[dashed] (-0.5, 0) -- (0, -0.5);
\draw[dashed] (0.5, 0) -- (0, -0.5);
\end{tikzpicture} = (-2\mathrm{i})^2 \intsinf \mathrm{d}s_1 \mathrm{d}s_2 \intinf \frac{\mathrm{d}p_1 \mathrm{d}p_2 \mathrm{d}q}{\pi^3} \frac{p_1 p_2}{(p_1^2 + (x_1 + s_1)^2)(p_2^2 + (x_2 + s_2)^2)} \frac{\delta(p_1 + q) \delta(p_2 - q)}{q^2 + y^2}
\end{eqn}
Note the similarity of this formula with \eqref{2siteflatmomspace}. The main difference is that the $x$'s in the numerators of the auxiliary propagators are now replaced by $p$. Hence, to obtain the time integral representation we use the relation 
\begin{eqn}
\mathrm{i} \sgn(t)e^{- x|t|} = \intinf \frac{\mathrm{d}p}{\pi} \frac{p}{p^2 + x^2} e^{\mathrm{i} p t }~.
\end{eqn}
Using this we can retrace the steps in \eqref{2siteflatmomspace} leading to the following time integral representation\footnote{The careful reader will notice the similarity of this equation with the 2-site wavefunction coefficient. Indeed, for this simple example the diagram with 2 dashed propagators returns the 2-site wavefunction coefficients and the equality can be proven via the following relation 
\begin{eqn}
&\intsinf \mathrm{d}s_1 \mathrm{d}s_2 \intinf \mathrm{d}t_1 \mathrm{d}t_2 \sgn(t_1) \sgn(t_2) e^{- (x_1 + s_1)|t_1|} e^{- (x_2 + s_2)|t_2|} e^{- y|t_1 - t_2|} \\
&= 2\intsinf \mathrm{d}s_1 \mathrm{d}s_2 \int_{-\infty}^0 \mathrm{d}t_1 \mathrm{d}t_2 e^{- (x_1 + s_1)t_1} e^{- (x_2 + s_2)t_2} \Big[ e^{- y|t_1 - t_2|} - e^{- y(t_1 + t_2)} \Big]
\end{eqn}
where the second line is the well known expression for the wavefunction coefficient of the 2-site graph \cite{Arkani-Hamed:2017fdk}. }
\begin{eqn}\label{2sitedashedtime1}
\begin{tikzpicture}[baseline]
\draw (-0.75, 0.5) -- (-0.5, 0);
\draw (-0.75, -0.5) -- (-0.5, 0);
\draw (-0.5, 0) -- (0.5, 0);
\draw (0.75, 0.5) -- (0.5, 0);
\draw (0.75, -0.5) -- (0.5, 0);
\draw[dashed] (-0.5, 0) -- (0, -0.5);
\draw[dashed] (0.5, 0) -- (0, -0.5);
\end{tikzpicture} 
= \intsinf \mathrm{d}s_1 \mathrm{d}s_2 \intinf \mathrm{d}t_1 \mathrm{d}t_2 \sgn(t_1) \sgn(t_2) e^{- (x_1 + s_1) |t_1|} e^{- (x_2 + s_2) |t_2|} G_F(t_1, t_2; y)
\end{eqn}
From this we obtain a time integral representation for the $s-$integrand. We diagrammatically denote these via
\begin{eqn}\label{blackdot1}
\begin{tikzpicture}[baseline]
\draw (-0.75, 0.5) -- (-0.5, 0);
\draw (-0.75, -0.5) -- (-0.5, 0);
\draw (-0.5, 0) -- (0.5, 0);
\draw (0.75, 0.5) -- (0.5, 0);
\draw (0.75, -0.5) -- (0.5, 0);
\draw[dashed] (-0.5, 0) -- (0, -0.5);
\draw[dashed] (0.5, 0) -- (0, -0.5);
\end{tikzpicture}  &= 
\intsinf \mathrm{d}s_1 \mathrm{d}s_2 \begin{tikzpicture}[baseline]
\draw (-0.5, 0) -- (0.5,0);
\node at (-0.5, 0) {$\bullet$};
\node at (0.5, 0) {$\bullet$};
\node at (-0.75, -0.25) {$x_1 + s_1$};
\node at (0.75, -0.25) {$x_2 + s_2$}; 
\node at (0, +0.25) {$y$};
\end{tikzpicture}, \\
\begin{tikzpicture}[baseline]
\draw (-0.5, 0) -- (0.5,0);
\node at (-0.5, 0) {$\bullet$};
\node at (0.5, 0) {$\bullet$};
\node at (-0.75, -0.25) {$X_1$};
\node at (0.75, -0.25) {$X_2$}; 
\node at (0, +0.25) {$y$};
\end{tikzpicture} &= \intinf \mathrm{d}t_1 \mathrm{d}t_2 \sgn(t_1) \sgn(t_2) e^{- X_1 |t_1|} e^{- X_2 |t_2|} G_F(t_1, t_2; y)
\end{eqn}
Thus the graphs with dashed propagators can be obtained by an $x-$shift of graphs formed by black vertices. The graphs with black vertices mimic the structure of \eqref{2sitefltatime1} with the main difference being the insertion of the $\sgn(t)$ function at each vertex. As discussed below \eqref{eq:dottedexfinal} these graphs are naturally interpreted as correlators of the conjugate momentum. 


For the dotted propagators we can assign a similar rule. In momentum-space they are simply constants $-\pi $ (see equation \eqref{eq:dotprop-review}) which implies in the time-representation they become $\delta(t)$ functions, 
\begin{eqn}
-\pi = -\pi\intinf \mathrm{d}t \delta(t)
\end{eqn}
Thus their contribution to the 2-site graph can be written as,
\begin{eqn}
\begin{tikzpicture}[baseline]
\draw (-0.75, 0.5) -- (-0.5, 0);
\draw (-0.75, -0.5) -- (-0.5, 0);
\draw (-0.5, 0) -- (0.5, 0);
\draw (0.75, 0.5) -- (0.5, 0);
\draw (0.75, -0.5) -- (0.5, 0);
\draw[dotted] (-0.5, 0) -- (0, -0.5);
\draw[dotted] (0.5, 0) -- (0, -0.5);
\end{tikzpicture}  = 
\pi^2 \intinf \mathrm{d}t_1 \mathrm{d}t_2 \delta(t_1) \delta(t_2) G_F(t_1, t_2; y)
\end{eqn}
Combining both propagators we obtain the time integral representation for the full 2-site tree level correlator 
\begin{eqn}
&\begin{tikzpicture}[baseline]
\draw (-0.75, 0.5) -- (-0.5, 0);
\draw (-0.75, -0.5) -- (-0.5, 0);
\draw (-0.5, 0) -- (0.5, 0);
\draw (0.75, 0.5) -- (0.5, 0);
\draw (0.75, -0.5) -- (0.5, 0);
\draw[dashed] (-0.5, 0) -- (0, -0.5);
\draw[dashed] (0.5, 0) -- (0, -0.5);
\end{tikzpicture} 
+ 
\begin{tikzpicture}[baseline]
\draw (-0.75, 0.5) -- (-0.5, 0);
\draw (-0.75, -0.5) -- (-0.5, 0);
\draw (-0.5, 0) -- (0.5, 0);
\draw (0.75, 0.5) -- (0.5, 0);
\draw (0.75, -0.5) -- (0.5, 0);
\draw[dotted] (-0.5, 0) -- (0, -0.5);
\draw[dotted] (0.5, 0) -- (0, -0.5);
\end{tikzpicture} \\
&= \intinf \mathrm{d}t_1 \mathrm{d}t_2 \Bigg[  \intsinf \mathrm{d}s_1 \mathrm{d}s_2 \sgn(t_1) \sgn(t_2) e^{- (x_1 + s_1) |t_1|} e^{- (x_2 + s_2)|t_2|} + \pi^2 \delta(t_1) \delta(t_2)  \Bigg] G_F(t_1, t_2; y) \\
&= \intinf \mathrm{d}t_1 \mathrm{d}t_2 \Bigg[ \frac{1}{t_1 t_2} +\pi^2 \delta(t_1) \delta(t_2)  \Bigg] e^{- x_1  |t_1|} e^{- x_2 |t_2|}  G_F(t_1, t_2; y) .
\end{eqn}
This gives general prescription for the time integral representation corresponding to the dressing rules \eqref{eq:dashprop-review} and \eqref{eq:dotprop-review} where at each site we have the effective bulk-boundary propagator,
\begin{eqn}
\phi_{\rm dS}(x, t) &\equiv\intsinf \mathrm{d}s \sgn(t) e^{- (x + s)|t|} - \pi \delta(t)  = \left[ \frac{1}{t} - \pi \delta(t) \right] e^{- x |t|}~.
\end{eqn}
The bulk-bulk propagators connecting each site are the standard Feynman propagators $G_F(t_1, t_2, y)$. It would be very interesting to derive these rules systematically by following the prescription of \cite{Donath:2024utn}. 

Since the dotted propagators arise via $\delta(t)$ at each vertex their contribution can be expressed in terms of a lower point diagram comprising of dashed propagators. For example consider the 3-site graph with one dotted propagator,
\begin{eqn}
\begin{tikzpicture}[baseline]
\draw (-0.75, 0.5) -- (-0.5, 0);
\draw (-0.75, -0.5) -- (-0.5, 0);
\draw (-0.5, 0) -- (0.5, 0);
\draw (0.75, 0.5) -- (0.5, 0);
\draw (0.75, -0.5) -- (0.5, 0);
\draw (0,0) -- (0,0.5);
\draw[dashed] (0, 0) -- (0, -0.5);
\draw[dashed] (-0.5, 0) -- (0, -0.5);
\draw[dotted] (0.5, 0) -- (0, -0.5);
\end{tikzpicture} 
&= -\pi \intinf \frac{\mathrm{d}t_1 \mathrm{d}t_2 \mathrm{d}t_3}{t_1 t_2} e^{- x_1 |t_1| - x_2 |t_2|} \delta(t_3) G_F(t_1, t_2; y_1) G_F(t_2, t_3; y_2)
\end{eqn}
By integrating over $t_3$ and using the relation $G(t, 0; y) = \frac{e^{- y |t|}}{y}$ we get
\begin{eqn}\label{eq:dottedexfinal}
\begin{tikzpicture}[baseline]
\draw (-0.75, 0.5) -- (-0.5, 0);
\draw (-0.75, -0.5) -- (-0.5, 0);
\draw (-0.5, 0) -- (0.5, 0);
\draw (0.75, 0.5) -- (0.5, 0);
\draw (0.75, -0.5) -- (0.5, 0);
\draw (0,0) -- (0,0.5);
\draw[dashed] (0, 0) -- (0, -0.5);
\draw[dashed] (-0.5, 0) -- (0, -0.5);
\draw[dotted] (0.5, 0) -- (0, -0.5);
\end{tikzpicture} 
&=  - \frac{\pi}{y_2} \intinf \frac{\mathrm{d}t_1 \mathrm{d}t_2}{t_1 t_2} e^{- x_1 |t_1| - (x_2 + y_2) |t_2|}  G_F(t_1, t_2; y_1) 
\end{eqn}
which is equivalent to the 2-site graph given in  \eqref{2sitedashedtime1} with shifted energies. More generally, for any dotted propagator within a graph, it factorizes it into two parts $\mathcal{I}_L$ and $\mathcal{I}_R$, with shifted energies on neighboring vertices:
\begin{equation}\label{eq:recusion_formula}
    {\begin{tikzpicture}[baseline,scale=0.7]
\draw[thick] (-1.5,0) circle (1);
\node at (-1.5,0) {$\mathcal{I}_L$};
\draw[thick] (1.5,0) circle (1);
\node at (1.5,0) {$\mathcal{I}_R$};
\draw[thick] (-.5,0)--(.5,0);
\node at (0,0) {$\bullet$};
\node[below] at (0,0) {$x_k$};
\draw[thick,gray] (-1.9,.9)--(0,2);
\draw[thick,gray] (-1.45,.97)--(0,2);
\node at (-1.,1.1) {.};
\node at (-.85,1.) {.};
\node at (-.7,.9) {.};
\draw[thick,gray] (-.7,.55)--(0,2);
\draw[thick,gray] (-.6,.4)--(0,2);
\draw[thick,gray] (1.9,.9)--(0,2);
\draw[thick,gray] (1.45,.97)--(0,2);
\node at (1.,1.1) {.};
\node at (.85,1.) {.};
\node at (.7,.9) {.};
\draw[thick,gray] (.7,.55)--(0,2);
\draw[thick,gray] (.6,.4)--(0,2);
\draw[thick,dotted] (0,0)--(0,2);
\node at (0,1.95) {\color{gray}\textbullet};
\end{tikzpicture}=\frac{-\pi}{y_{k-1}y_k}\times\begin{tikzpicture}[baseline,scale=0.7]
\draw[thick] (0,0) circle (1);
\node at (0,0) {$\mathcal{I}_L$};
\node at (0,1.95) {\color{gray}\textbullet};
\draw[gray,thick] (-.71,.71)--(0,2);
\draw[gray,thick] (-.5,.865)--(0,2);
\draw[gray,thick] (.71,.71)--(0,2);
\draw[gray,thick] (.5,.865)--(0,2);
\node at (0,1.15) {\scriptsize$\dots$};
\node[right] at (1,0) {\scriptsize$x_{k-1}+y_{k-1}$};
\node at (1,0) {$\bullet$};
\end{tikzpicture}\times\begin{tikzpicture}[baseline,scale=0.7]
\draw[thick] (0,0) circle (1);
\node at (0,0) {$\mathcal{I}_R$};
\node at (0,1.95) {\color{gray}\textbullet};
\draw[gray,thick] (-.71,.71)--(0,2);
\draw[gray,thick] (-.5,.865)--(0,2);
\draw[gray,thick] (.71,.71)--(0,2);
\draw[gray,thick] (.5,.865)--(0,2);
\node at (0,1.15) {\scriptsize$\dots$};
\node[left] at (-1,0) {\scriptsize$x_{k+1}+y_{k}$};
\node at (-1,0) {$\bullet$};
\end{tikzpicture}},
\end{equation}
in which the gray propagators indicate that the corresponding auxiliary lines are not specified to be either dotted or dashed.
We revisit diagrams with dotted propagators in section \ref{sec:simplicity}. 
While graphs with white vertices admit the usual interpretation as flat space correlators of fields, graphs with black vertices naturally compute correlators with external conjugate momentum insertions.
This follows directly from the perturbative path integral for time-ordered correlators.
The standard bulk-boundary propagator for a field arises from the contraction $\braket{\phi(0) \phi(t)}=\frac{1}{\omega}e^{- \omega|t|}$, whereas inserting the conjugate momentum on an external leg gives
\[
\braket{\pi(0) \phi(t)}= \lim_{t' \to 0} \p_{t'}\braket{\phi(t') \phi(t)}
= \left. \frac{1}{\omega} \p_{t'} e^{- \omega |t - t'|} \right|_{t' = 0}
= \sgn(t) e^{- \omega|t|}.
\]
Thus the de Sitter correlators of fields are obtained by integrating over energies of flat space correlators with conjugate momentum insertions.
This is an important conceptual difference from the wavefunction story.
Wavefunction coefficients in de Sitter space are obtained by integrating flat space wavefunction coefficients in the same field basis, whereas for correlators the natural flat space precursor lives in a mixed field/momentum basis.
In this form the representation is both direct and flexible, and it will provide the basic language for all subsequent recursion relations and expansions.
The main output of this section is therefore not only a new way of writing the same correlator, but also a change of basis in which the operations relevant to the correlator close naturally. We now turn to the consequences of this change of basis: parity constraints, recursive reductions, melonic collapses, and singularity expansions all become simple statements about the black/white time integral graphs.


\section{The Simplicity of dS Correlators}\label{sec:simplicity}

In this section we use the time integral representation derived above to expose a number of simplicity properties of de Sitter correlators.
The guiding idea is that the apparent complexity of the dressing rules is reorganized once the correlator is viewed as part of a larger family of mixed field/momentum correlators.
It is convenient to work with graphs containing both black and white vertices, representing conjugate momentum and field insertions respectively.
Our final object of interest is the graph with all vertices black, since integrating over its shifted energies gives the de Sitter correlator, but the broader class is closed under the recursive operations we derive below.
The time integral representation for such a generic graph is
\begin{eqn}\label{eq:time_representation}
\braket{G}_{B;W}=\intinf \prod_{i=1}^B \mathrm{d}t_i \sgn(t_i)e^{-x_i|t_i|} \prod_{j=B+1}^{W} \mathrm{d}t_j e^{-x_j |t_j|} \prod_{e=1}^EG_F(t,t';y_e),
\end{eqn}
where $\{t, t'\}\in\{t_1,\dots,t_B, t_{B+1}, \cdots, t_{B+W}\}$.

\subsection{Vanishing Contributions and Weight Drop}\label{ssec:oddchain}
The first simplification is immediate: any graph with an odd number of black vertices vanishes.
Indeed, under the change of variables $t_i \to - t_i$ for all $i\in\{1,\cdots,B+W\}$, the integrand in \eqref{eq:time_representation} acquires a factor of $(-1)^B$ from the sign insertions.
Hence, for odd $B=2n+1$, \eqref{eq:time_representation} becomes
\begin{eqn}
\braket{G}_{2n+1; W}=(-1)\intinf \prod_{i=1}^{2n+1} \mathrm{d}t_i \sgn(t_i)e^{-x_i|t_i|} \prod_{j=B+1}^{W} \mathrm{d}t_j e^{-x_j|t_j|} \prod_{e=1}^EG_F(t,t';y_e)
\end{eqn}
implying that it vanishes, $\braket{G}_{2n+1; W} = 0$.
This also gives a simple explanation of the weight drop for odd-point
de Sitter correlators. The maximal-weight sector is the all-dashed
sector, which contains one energy-integral over $s$ for each vertex. For odd $n$,
however, this sector is an odd-black-vertex graph and hence vanishes. The first
non-vanishing contribution must contain one dotted propagator. By
\eqref{eq:recusion_formula}, this dotted insertion collapses one vertex and
reduces the answer to a corresponding $(n-1)$-black-vertex graph, with shifted
neighboring energies and simple rational edge factors, multiplied by an overall
factor of $-\pi$. Thus the highest-weight part of an odd-point correlator behaves
like the corresponding $(n-1)$-point object, and its maximal weight is reduced
from $n$ to $n-1$.

We now derive several recursion relations for graphs built from black $\bullet$ and white $\circw$ vertices.
To streamline the presentation, it is convenient to introduce the notation
\begin{eqn}\label{proprec1}
\xi(x, t) = e^{- x |t|}, \qquad  
\phi(x, t) =  \sgn(t)e^{- x |t|}.
\end{eqn}
where these denote the bulk-boundary propagators for the white and black vertices respectively.

\subsection{Leaf Recursion Relations}
In this subsection, we derive recursion relations for graphs with a leaf, namely a single vertex attached to the rest of the graph by a single edge.
These identities are local: whenever a graph contains a leaf, the time integral over this leaf can be performed independently of the remaining subgraph.
Thus the relations below are not restricted to tree graphs, although they become especially powerful there.
Indeed, any higher-point tree has at least one leaf, so repeated use of the same local recursion reduces the tree step by step to lower-point building blocks.

We start by noticing that the propagators introduced in \eqref{proprec1} satisfy the following integration relations:
\begin{eqn}\label{rec1}
\intinf \mathrm{d}t_1 \phi(x, t_1) G_F(t_1, t_2, y) &= \frac{2}{(y^2 - x^2)} \Big[ \phi(x, t_2) - \phi(y, t_2)  \Big], \\
\intinf \mathrm{d}t_1 \xi(x, t_1) G_F(t_1, t_2, y) &= \frac{2}{(y^2 - x^2)} \Big[  \xi(x, t_2) - \frac{x}{y} \xi(y, t_2)  \Big],
\end{eqn}
For example consider a generic graph $F$ connecting with a leaf $x_1$: 
\begin{eqn}
\begin{tikzpicture}[baseline]
\node at (-1, 0) {$\bullet$};
\node at (0,0) {$\bullet$};
\draw (0.55, 0) circle (0.5);
\draw (-1, 0) -- (0,0);
\node at (0.5, 0) {$F$};
\node at (-1, -0.25) {$x_1$};
\node at (0, -0.25) {$x_2$};
\node at (-0.5, +0.25) {$y$};
\end{tikzpicture} = \intinf \mathrm{d}t_1 \phi(x_1, t_1) G_F(t_1, t_2, y) \phi(x_2, t_2) F(t_2).
\end{eqn}
 Using \eqref{rec1} we obtain 
\begin{eqn}
\begin{tikzpicture}[baseline]
\node at (-1, 0) {$\bullet$};
\node at (0,0) {$\bullet$};
\draw (0.55, 0) circle (0.5);
\draw (-1, 0) -- (0,0);
\node at (0.5, 0) {$F$};
\node at (-1, -0.25) {$x_1$};
\node at (0, -0.25) {$x_2$};
\node at (-0.5, +0.25) {$y$};
\end{tikzpicture} &= \frac{2}{(y^2 - x_1^2)} \intinf  \mathrm{d}t_2 \Big[ \sgn(t_2)\phi(x_1 + x_2, t_2) - \sgn(t_2) \phi(x_2 + y, t_2) \Big]F(t_2) \\
&= \frac{2}{(y^2 - x_1^2)}
\Bigg[
\begin{tikzpicture}[baseline]
\draw (0.55, 0) circle (0.5);
\node at (0,0) {$\circw$};
\node at (-0.5, -0.25) {$x_1 + x_2$};
\node at (0.5, 0) {$F$};
\end{tikzpicture}
- 
\begin{tikzpicture}[baseline]
\draw (0.55, 0) circle (0.5);
\node at (0,0) {$\circw$};
\node at (-0.5, -0.25) {$x_2 + y$};
\node at (0.5, 0) {$F$};
\end{tikzpicture}
\Bigg]~.
\end{eqn}
These relations can be generalized for all possible combinations of black and white vertices. We summarize all such recursions below, 
\begin{eqn}\label{treerec1}
\begin{tikzpicture}[baseline]
\node at (-1, 0) {$\bullet$};
\node at (0,0) {$\bullet$};
\draw (0.55, 0) circle (0.5);
\draw (-1, 0) -- (0,0);
\node at (0.5, 0) {$F$};
\node at (-1, -0.25) {$x_1$};
\node at (0, -0.25) {$x_2$};
\node at (-0.5, +0.25) {$y$};
\end{tikzpicture}
&= \frac{2}{(y^2 - x_1^2)}
\Bigg[
\begin{tikzpicture}[baseline]
\draw (0.55, 0) circle (0.5);
\node at (0,0) {$\circw$};
\node at (-0.5, -0.25) {$x_1 + x_2$};
\node at (0.5, 0) {$F$};
\end{tikzpicture}
- 
\begin{tikzpicture}[baseline]
\draw (0.55, 0) circle (0.5);
\node at (0,0) {$\circw$};
\node at (-0.5, -0.25) {$x_2 + y$};
\node at (0.5, 0) {$F$};
\end{tikzpicture}
\Bigg], \\
\begin{tikzpicture}[baseline]
\node at (-1, 0) {$\bullet$};
\draw (0.55, 0) circle (0.5);
\draw (-1, 0) -- (0,0);
\node at (0,0) {$\circw$};
\node at (0.5, 0) {$F$};
\node at (-1, -0.25) {$x_1$};
\node at (0, -0.25) {$x_2$};
\node at (-0.5, +0.25) {$y$};
\end{tikzpicture}
&= \frac{2}{(y^2 - x_1^2)}
\Bigg[
\begin{tikzpicture}[baseline]
\node at (0,0) {$\bullet$};
\draw (0.55, 0) circle (0.5);
\node at (-0.5, -0.25) {$x_1 + x_2$};
\node at (0.5, 0) {$F$};
\end{tikzpicture}
- 
\begin{tikzpicture}[baseline]
\node at (0,0) {$\bullet$};
\draw (0.55, 0) circle (0.5);
\node at (-0.5, -0.25) {$x_2 + y$};
\node at (0.5, 0) {$F$};
\end{tikzpicture}
\Bigg], \\
\begin{tikzpicture}[baseline]
\draw (0.55, 0) circle (0.5);
\draw (-1, 0) -- (0,0);
\node at (-1, 0) {$\circw$};
\node at (0,0) {$\circw$};
\node at (0.5, 0) {$F$};
\node at (-1, -0.25) {$x_1$};
\node at (0, -0.25) {$x_2$};
\node at (-0.5, +0.25) {$y$};
\end{tikzpicture}
&= \frac{2}{(y^2 - x_1^2)}
\Bigg[
\begin{tikzpicture}[baseline]
\draw (0.55, 0) circle (0.5);
\node at (0,0) {$\circw$};
\node at (-0.5, -0.25) {$x_1 + x_2$};
\node at (0.5, 0) {$F$};
\end{tikzpicture}
- \frac{x_1}{y}
\begin{tikzpicture}[baseline]
\draw (0.55, 0) circle (0.5);
\node at (0,0) {$\circw$};
\node at (-0.5, -0.25) {$x_2 + y$};
\node at (0.5, 0) {$F$};
\end{tikzpicture}
\Bigg],\\
\begin{tikzpicture}[baseline]
\node at (0,0) {$\bullet$};
\draw (0.55, 0) circle (0.5);
\draw (-1, 0) -- (0,0);
\node at (-1, 0) {$\circw$};
\node at (0.5, 0) {$F$};
\node at (-1, -0.25) {$x_1$};
\node at (0, -0.25) {$x_2$};
\node at (-0.5, +0.25) {$y$};
\end{tikzpicture}
&= \frac{2}{(y^2 - x_1^2)}
\Bigg[
\begin{tikzpicture}[baseline]
\node at (0,0) {$\bullet$};
\draw (0.55, 0) circle (0.5);
\node at (-0.5, -0.25) {$x_1 + x_2$};
\node at (0.5, 0) {$F$};
\end{tikzpicture}
- \frac{x_1}{y}
\begin{tikzpicture}[baseline]
\node at (0,0) {$\bullet$};
\draw (0.55, 0) circle (0.5);
\node at (-0.5, -0.25) {$x_2 + y$};
\node at (0.5, 0) {$F$};
\end{tikzpicture}
\Bigg]. 
\end{eqn}
 In particular, the third equation, which denotes the recursion relations for the flat space correlator, was also given in \cite{Donath:2024utn}. We remark that the recursion relations \eqref{treerec1} are especially useful at tree level. As an example we can use these set of relations to derive the 4-site chain by starting with the 2-site chain (which can further be obtained in terms of the single site graph), 
\begin{eqn}\label{chain_recursion}
\begin{tikzpicture}[baseline]
\node at (-1, 0) {$\bullet$};
\node at (1, 0) {$\bullet$};
\draw (-1, 0) -- (1, 0);
\node at (-1, -0.25) {$x_1$};
\node at (1, -0.25) {$x_2$};
\node at (0, 0.25) {$y$};
\end{tikzpicture}
&= \frac{4}{(x_1 + x_2)(x_1 + y)(x_2 + y)}, \\
\begin{tikzpicture}[baseline]
\node at (0, 0) {$\bullet$};
\node at (1, 0) {$\bullet$};
\node at (-1, -0.25) {$\tilde x$};
\node at (0, -0.25) {$x_3$};
\node at (1, -0.25) {$x_4$};
\node at (-0.5, 0.25) {$y_2$};
\node at (0.5, 0.25) {$y_3$};
\draw (-1, 0) -- (1,0);
\node at (-1, 0) {$\circw$};
\end{tikzpicture}
&= \frac{2}{y_2^2 - \tilde x^2} \Big[ 
\begin{tikzpicture}[baseline]
\node at (-1, 0) {$\bullet$};
\node at (1, 0) {$\bullet$};
\draw (-1, 0) -- (1, 0);
\node at (-1, -0.25) {$x_3 + \tilde x$};
\node at (1, -0.25) {$x_4$};
\node at (0, 0.25) {$y_3$};
\end{tikzpicture}
- \frac{\tilde x}{y_2}
\begin{tikzpicture}[baseline]
\node at (-1, 0) {$\bullet$};
\node at (1, 0) {$\bullet$};
\draw (-1, 0) -- (1, 0);
\node at (-1, -0.25) {$x_3 + y_2$};
\node at (1, -0.25) {$x_4$};
\node at (0, 0.25) {$y_3$};
\end{tikzpicture}
\Big], \\
\begin{tikzpicture}[baseline]
\draw (-1.5, 0) -- (1.5,0);
\node at (-1.5, 0) {$\bullet$};
\node at (-0.5, 0) {$\bullet$};
\node at (0.5, 0) {$\bullet$};
\node at (1.5, 0) {$\bullet$};
\node at (-1.5, -0.25) {$x_1$};
\node at (-0.5, -0.25) {$x_2$};
\node at (1.5, -0.25) {$x_4$};
\node at (0.5, -0.25) {$x_3$};
\node at (-1, 0.25) {$y_1$};
\node at (0, 0.25) {$y_2$};
\node at (1.25, 0.25) {$y_3$};
\end{tikzpicture}
&= \frac{2}{y_1^2 - x_1^2} \Big[ 
\begin{tikzpicture}[baseline]
\node at (0, 0) {$\bullet$};
\node at (1, 0) {$\bullet$};
\node at (-1, -0.25) {$x_2 + x_1$};
\node at (0, -0.25) {$x_3$};
\node at (1, -0.25) {$x_4$};
\node at (-0.5, 0.25) {$y_2$};
\node at (0.5, 0.25) {$y_3$};
\draw (-1, 0) -- (1,0);
\node at (-1, 0) {$\circw$};
\end{tikzpicture}
- 
\begin{tikzpicture}[baseline]
\node at (0, 0) {$\bullet$};
\node at (1, 0) {$\bullet$};
\node at (-1, -0.25) {$x_2 + y_1$};
\node at (0, -0.25) {$x_3$};
\node at (1, -0.25) {$x_4$};
\node at (-0.5, 0.25) {$y_2$};
\node at (0.5, 0.25) {$y_3$};
\draw (-1, 0) -- (1,0);
\node at (-1, 0) {$\circw$};
\end{tikzpicture}
\Big]
\end{eqn}
Solving this recursion gives, 
\begin{eqn}\label{chainexample1}
&\begin{tikzpicture}[baseline]
\draw (-1.5, 0) -- (1.5,0);
\node at (-1.5, 0) {$\bullet$};
\node at (-0.5, 0) {$\bullet$};
\node at (0.5, 0) {$\bullet$};
\node at (1.5, 0) {$\bullet$};
\node at (-1.5, -0.25) {$x_1$};
\node at (-0.5, -0.25) {$x_2$};
\node at (1.5, -0.25) {$x_4$};
\node at (0.5, -0.25) {$x_3$};
\node at (-1., 0.25) {$y_1$};
\node at (0, 0.25) {$y_2$};
\node at (1.25, 0.25) {$y_3$};
\end{tikzpicture}=16\times\\
& \frac{\frac{\frac{1}{\left(x_1+x_2+x_3+x_4\right) \left(x_1+x_2+x_3+y_3\right)}-\frac{x_1+x_2}{y_2 \left(x_3+x_4+y_2\right) \left(x_3+y_2+y_3\right)}}{y_2^2-\left(x_1+x_2\right){}^2}+\frac{\frac{1}{\left(x_2+x_3+x_4+y_2\right) \left(x_2+x_3+y_2+y_3\right)}-\frac{x_2+y_2}{y_2 \left(x_3+x_4+y_2\right) \left(x_3+y_2+y_3\right)}}{x_2 \left(x_2+2 y_2\right)}}{\left(y_2^2-x_1^2\right) \left(x_4+y_3\right)}.
\end{eqn}

We can also use this recursion to solve for tree level graphs of other topologies, for example, star-graphs: 
\begin{eqn}
\begin{tikzpicture}[baseline]
\draw (0, 1) -- (0,0);
\draw (-0.7, -0.7) -- (0,0);
\draw (0.7, -0.7) -- (0,0);
\node at (0, 0) {$\bullet$};
\node at (0.7, -0.7) {$\bullet$};
\node at (-0.7, -0.7) {$\bullet$};
\node at (0, 1) {$\bullet$};
\node at (0, 1.25) {$x_1$};
\node at (-0.85, -0.85) {$x_2$};
\node at (0.85, -0.85) {$x_3$};
\node at (0, -0.25) {$x_4$};
\node at (-0.25, 0.5) {$y_{14}$};
\end{tikzpicture}
&= \frac{2}{y_{14}^2 - x_1^2} \Big[ \begin{tikzpicture}[baseline]
\node at (-1, 0) {$\bullet$};
\node at (1, 0) {$\bullet$};
\node at (-1.25, -0.25) {$x_2$};
\node at (0, -0.25) {$x_4 + x_1$};
\node at (1.25, -0.25) {$x_3$};
\node at (-0.5, 0.25) {$y_{24}$};
\node at (0.5, 0.25) {$y_{34}$};
\draw (-1, 0) -- (1,0);
\node at (0, 0) {$\circw$};
\end{tikzpicture}
- 
\begin{tikzpicture}[baseline]
\node at (-1, 0) {$\bullet$};
\node at (1, 0) {$\bullet$};
\node at (-1.25, -0.25) {$x_2$};
\node at (0, -0.25) {$x_4 + y_{14}$};
\node at (1.25, -0.25) {$x_3$};
\node at (-0.5, 0.25) {$y_{24}$};
\node at (0.5, 0.25) {$y_{34}$};
\draw (-1, 0) -- (1,0);
\node at (0, 0) {$\circw$};
\end{tikzpicture}
\Big], \\
\begin{tikzpicture}[baseline]
\node at (-1, 0) {$\bullet$};
\node at (1, 0) {$\bullet$};
\node at (-1.25, -0.25) {$x_1$};
\node at (0, -0.25) {$x_2$};
\node at (1.25, -0.25) {$x_3$};
\node at (-0.5, 0.25) {$y_1$};
\node at (0.5, 0.25) {$y_2$};
\draw (-1, 0) -- (1,0);
\node at (0, 0) {$\circw$};
\end{tikzpicture} &= \frac{2}{y_1^2 - x_1^2} \Big[ 
\begin{tikzpicture}[baseline]
\node at (-1, 0) {$\bullet$};
\node at (1, 0) {$\bullet$};
\draw (-1, 0) -- (1, 0);
\node at (-1, -0.25) {$x_2 + x_1$};
\node at (1, -0.25) {$x_3$};
\node at (0, 0.25) {$y_2$};
\end{tikzpicture} 
- \begin{tikzpicture}[baseline]
\node at (-1, 0) {$\bullet$};
\node at (1, 0) {$\bullet$};
\draw (-1, 0) -- (1, 0);
\node at (-1, -0.25) {$x_2 + y_1$};
\node at (1, -0.25) {$x_3$};
\node at (0, 0.25) {$y_2$};
\end{tikzpicture}
\Big]~.
\end{eqn}
These relations allow us to generate arbitrary tree graphs recursively.
Combined with the melonic reductions discussed later in Section~\ref{sssec:melon}, they also provide efficient control over a broad class of loop integrands built from trees and melons.

\subsection*{Momentum-space Recursion}
The same recursion relations admit a clean momentum-space derivation, which is also convenient for obtaining closed-form expressions.
To proceed, we denote the Fourier transforms of the propagators in \eqref{proprec1} by
\begin{eqn}
     \phi(x,t)=\int_{-\infty}^{\infty} \frac{\mathrm{d}p}{2\pi} E_x(p)e^{ \mathrm{i}pt},\qquad\xi(x,t)=\int_{-\infty}^{\infty}\frac{ \mathrm{d}p}{2\pi} H_x(p)e^{\mathrm{i}pt},
\end{eqn}
where
\begin{equation}
    E_x(q)=\frac{-2\mathrm{i}q}{q^2+x^2},\qquad\text{and}\qquad H_x(q)=\frac{2x}{q^2+x^2}
\end{equation}
to denote the counterparts of black dots $\bullet$ and white dots $\circ$ in momentum-space respectively. We also require the Fourier space representation of the Feynman propagator given in \eqref{eq:feynprop1}. We now consider the Fourier space representation of a  graph with $E$-edges containing $B$-black and $W$-vertices whose time integral representation is given in \eqref{eq:time_representation},
\begin{equation}\label{eq:GBWgeneral1}
    \begin{aligned}
	\langle G\rangle _{B;W}&=\int_{-\infty}^{\infty}{\prod_{i=1}^B{\mathrm{d}t_i\frac{\mathrm{d}p_i}{2\pi}E_{x_i}}(p_i)e^{\mathrm{i}p_it_i}\prod_{j=B+1}^{B+W}{\mathrm{d}t_j\frac{\mathrm{d}p_j}{2\pi}H_{x_j}}(p_j)e^{\mathrm{i}p_jt_j}\prod_{e=1}^E{\frac{\mathrm{d}q_e}{\pi}\frac{e^{\mathrm{i}q_e(t-t ^\prime)}}{q_{e}^{2}+y_{e}^{2}}}}\\
	&=\int_{-\infty}^{\infty}{\prod_{i=1}^B{\mathrm{d}p_iE_{x_i}}(p_i)\delta \left( p_i-\mathcal{P} _i(q) \right) \prod_{j=B+1}^{B+W}{\mathrm{d}p_jH_{x_j}}(p_j)\delta \left( p_j-\mathcal{P} _j(q) \right) \prod_{e=1}^E{\frac{\mathrm{d}q_e}{\pi}\frac{1}{q_{e}^{2}+y_{e}^{2}}}},\\
\end{aligned}
\end{equation}
in which $\{t, t'\}\in\{t_1,\dots,t_B, t_{B+1}, \cdots, t_{B+W}\}$, $\mathcal P_i(q)$ and $\mathcal{P}_j(q)$ are the polynomials of $q$'s determined by the graph structure and we provide explicit expressions for special cases below. We can now integrate out all $p$'s by using the delta functions arising from the time integral, obtaining an expression of the form,
\begin{equation}\label{eq:momentum_representation}
\langle G\rangle _{B;W}=\int_{-\infty}^{\infty}{\prod_{i=1}^B{E}_{x_i}(\mathcal{P} _i(q))\prod_{j=B+1}^{B+W}{H_{x_j}}(\mathcal{P} _j(q))\prod_{e=1}^E{\frac{\mathrm{d}q_e}{\pi}\frac{1}{q_{e}^{2}+y_{e}^{2}}}}.
\end{equation}
An advantage of performing recursions in momentum-space is that we can evaluate integrals without splitting them into different intervals, since there are no sign or absolute-value functions as in the time representation. 

It is easy to recover the result in section \ref{ssec:oddchain} by a simple change of variables ($q_i \to -q_i$) that \eqref{eq:nchainblack} and \eqref{eq:npolygonblack} vanish for odd $n$. To obtain the other recursions one can explicitly check that both $E_x(p)$ and $H_x(p)$ satisfies these algebraic equalities 
\begin{equation}\label{eq:algebraic_properties_by_EH}
    \frac{E_x(p)}{p^2+y^2}=\frac{E_y(p)-E_x(p)}{x^2-y^2},\qquad\frac{H_x(p)}{p^2+y^2}=\frac{\frac{x}{y}H_y(p)-H_x(p)}{x^2-y^2},
\end{equation}
which coincide with \eqref{rec1} after a Fourier transform and is the basis of the recursion relations in momentum-space. 
As an example we show how the recursion relations \eqref{treerec1} follows from the momentum-space representation. Consider a black vertex connected to a general graph, whose momentum-space representation is given as, 
\begin{equation}
\begin{tikzpicture}[baseline]
\node at (-1, 0) {$\bullet$};
\node at (0,0) {$\bullet$};
\draw (0.55, 0) circle (0.5);
\draw (-1, 0) -- (0,0);
\node at (0.5, 0) {$F$};
\node at (-1, -0.25) {$x_1$};
\node at (0, -0.25) {$x_2$};
\node at (-0.5, +0.25) {$y$};
\end{tikzpicture} = \int_{-\infty}^\infty \frac{\mathrm{d}q}{\pi} \frac{\mathrm{d}q'}{\pi} E_{x_1}(q)\frac{1}{q^2+y^2}E_{x_2}(-q+c(q'))F(q'),
\end{equation}
where $c(q')$ is the sum of energies flowing into the black vertex from the blob $F$ excluding $q$. By using the algebraic equality above, we can rewrite the right hand side as
\begin{equation}
\begin{aligned}
    \begin{tikzpicture}[baseline]
\node at (-1, 0) {$\bullet$};
\node at (0,0) {$\bullet$};
\draw (0.55, 0) circle (0.5);
\draw (-1, 0) -- (0,0);
\node at (0.5, 0) {$F$};
\node at (-1, -0.25) {$x_1$};
\node at (0, -0.25) {$x_2$};
\node at (-0.5, +0.25) {$y$};
    \end{tikzpicture}
&=\frac{1}{x_1^2-y^2}\int_{-\infty}^\infty \frac{\mathrm{d}q}{\pi} \frac{\mathrm{d}q'}{\pi}  \left[E_y(q)-E_{x_1}(q)\right]E_{x_2}(-q+c(q'))F(q')\\
&=\intinf \frac{\mathrm{d}q'}{\pi}  \frac{2F(q')}{y^2-x_1^2}[H_{x_1+x_2}(c)-H_{x_2+y}(c)]\\
&= \frac{2}{y^2-x_1^2}\left[\begin{tikzpicture}[baseline]
\node at (0,0) {$\circ$};
\draw (0.55, 0) circle (0.5);
\node at (0.5, 0) {$F$};
\node at (-.5, -0.25) {$x_1+x_2$};
    \end{tikzpicture}-\begin{tikzpicture}[baseline]
\node at (0,0) {$\circ$};
\draw (0.55, 0) circle (0.5);
\node at (0.5, 0) {$F$};
\node at (-.5, -0.25) {$x_2+y$};
    \end{tikzpicture}\right],
    \end{aligned}
\end{equation}
in which we have used the fact that
\begin{equation}
    \int_{-\infty}^\infty \frac{\mathrm{d}q}{\pi}E_{x_1}(q)E_{x_2}(-q+c)=2H_{x_1+x_2}(c).
\end{equation} 
By the analogous procedure, we can also derive the recursion relation for a white vertex, which leads us to rediscover the same recursion relations as in time space shown in \eqref{rec1}.

As an application, we can explicitly write down the momentum-space representation for $n$-site chain and $n$-site polygon as follows.
 For $G$ being an $n$-site chain, 
\begin{equation}
    \mathcal{P}_v(q)=\begin{cases}q_1, & v=1, \\ q_v-q_{v-1}, & 2\le v\le n-1, \\ -q_{n-1}, & v=n,\end{cases}
\end{equation}
thus with all vertices being black, we have
\begin{equation}\label{eq:nchainblack}
    \langle\text{$n$-chain}\rangle_{n;0}=\int_{-\infty}^\infty \frac{-2\mathrm{i}q_1}{q_1^2+x_1^2}\frac{2\mathrm{i}q_{n-1}}{q_{n-1}^2+x_n^2}\prod_{i=2}^{n-1}\frac{-2\mathrm i(q_i-q_{i-1})}{(q_i-q_{i-1})^2+x_i^2}\prod_{e=1}^{n-1}\frac{\mathrm{d}q_e}{\pi}\frac{1}{q_e^2+y_e^2},
\end{equation}
and with all vertices being white,
\begin{equation}
    \langle\text{$n$-chain}\rangle_{0;n}=\int_{-\infty}^\infty \frac{2 x_1}{q_1^2+x_1^2}\frac{2x_n}{q_{n-1}^2+x_n^2}\prod_{j=2}^{n-1}\frac{2x_j}{(q_j-q_{j-1})^2+x_j^2}\prod_{e=1}^{n-1}\frac{\mathrm{d}q_e}{\pi}\frac{1}{q_e^2+y_e^2}.
\end{equation}
As a direct implementation of these recursion relations, we can compute the momentum-space representation for any tree graph recursively. For example, for the $n$-black-dot-chain ($n$ being even), the result is given by
\begin{equation}
 \langle\text{$n$-chain}\rangle_{n;0}
=
\frac{1}{2^{n-4}}
\sum_{a=1}^{n-1}
\frac{c_{n-1,a}}{w_{a,n-1}^2-y_{n-1}^2}
\left(
\frac{1}{x_n+y_{n-1}}-\frac{1}{x_n+w_{a,n-1}}
\right),
\end{equation}
where the coefficients $c_{n-1,a}$ and $w_{a,n-1}$ are defined by the following:
\begin{align}
\label{eq:tubing}
w_{a,j}=
\begin{cases}
x_1+\cdots+x_j, & a=1,\\[4pt]
x_a+\cdots+x_j+y_{a-1}, & 2\le a\le j,
\end{cases}
\qquad\text{for } 1\le a\le j,
\end{align}
in this form, $w_{a,j}$ is naturally identified with the tubing for the consecutive tube enclosing the vertices $a,\ldots,j$; for $a>1$ the additional $y_{a-1}$ records the edge crossed by the left boundary of the tube, and
\begin{equation}
\begin{aligned}
c_{j,a}=\frac{4\left( -1 \right) ^j}{w_{a,j-1}^{2}-y_{j-1}^{2}}\,c_{j-1,a}\qquad\text{for } 1\le a\le j-1,\\c_{j,j}=
\begin{cases}
\displaystyle -\sum_{a=1}^{j-1}\frac{w_{a,j-1}}{y_{j-1}}\,c_{j,a},
& j>1\ \text{odd},\\[10pt]
\displaystyle -\sum_{a=1}^{j-1}c_{j,a},
& j>1\ \text{even},
\end{cases}\qquad\text{and } \qquad c_{1,1}=1.
\end{aligned}
\end{equation}

For example, when $n=2$
\begin{equation}
    w_{1,1}=x_1,
\qquad
c_{1,1}=1,
\end{equation}
so
\begin{equation}\label{eq:2chain}
\langle\text{$2$-chain}\rangle_{2;0}
=
4\,
\frac{1}{x_1^2-y^2}
\left(
\frac{1}{x_2+y}-\frac{1}{x_2+x_1}
\right)
=
\frac{4}{(x_1+x_2)(x_1+y)(x_2+y)},
\end{equation}
which agrees with the first equation of \eqref{chain_recursion}.

By the analogous procedure one derives the recursion relation for a white vertex as well, recovering exactly the same identities as in time space.
The two descriptions are therefore complementary: time space makes the geometric origin of the relations transparent, while momentum-space is often better suited for explicit evaluation.

\subsection{Fusion Recursions}
We can derive a second class of recursion relations by fusing lower-point graphs into higher-point trees or loop integrands.
Relations of this kind were previously explored for the wavefunction \cite{Benincasa:2018ssx, Benincasa:2024leu, Chowdhury:2024snc}; here we show how they generalize to correlators.


We first state the result for obtaining graphs with black vertices by fusing lower point graphs,
\begin{eqn}\label{eq:black_dot_fusion}
&\int_{-i\infty}^{+i\infty} \frac{\mathrm{d}w}{4\pi \mathrm{i}}  \quad 
\begin{tikzpicture}[baseline]
\draw (-1.5, 0) circle (0.5);
\node at (-1.5, 0) {$F_1$};
\draw (-1, 0) -- (0,0);
\node at (0, 0) {$\circw$};
\node at (0, -0.25) {\scriptsize $x_1+ w$};
\node at (-0.5, 0.25) {$y_1$};
\end{tikzpicture}
\begin{tikzpicture}[baseline]
\draw (1.5, 0) circle (0.5);
\node at (1.5, 0) {$F_2$};
\node at (0, 0) {$\bullet$};
\draw (1, 0) -- (0,0);
\node at (0, -0.25) {\scriptsize $x_2 - w$};
\node at (0.5, 0.25) {$y_2$};
\end{tikzpicture}
+ 
\begin{tikzpicture}[baseline]
\draw (-1.5, 0) circle (0.5);
\node at (-1.5, 0) {$F_1$};
\node at (0, 0) {$\bullet$};
\draw (-1, 0) -- (0,0);
\node at (0, -0.25) {\scriptsize $x_1+ w$};
\node at (-0.5, 0.25) {$y_1$};
\end{tikzpicture}
\begin{tikzpicture}[baseline]
\draw (1.5, 0) circle (0.5);
\node at (1.5, 0) {$F_2$};
\draw (1, 0) -- (0,0);
\node at (0, 0) {$\circw$};
\node at (0, -0.25) {\scriptsize $x_2 - w$};
\node at (0.5, 0.25) {$y_2$};
\end{tikzpicture}\\
&= 
\begin{tikzpicture}[baseline]
\draw (-1.5, 0) circle (0.5);
\draw (1.5, 0) circle (0.5);
\draw (-1, 0) -- (1,0);
\node at (0,0) {$\bullet$};
\node at (-1.5, 0) {$F_1$};
\node at (1.5, 0) {$F_2$};
\node at (-0.5, 0.25) {$y_1$};
\node at (0.5, 0.25) {$y_2$};
\node at (0, -0.25) {\scriptsize $x_1 + x_2$};
\end{tikzpicture}
\end{eqn}
This can be proven as follows. Consider the sum of two terms in the LHS,
\begin{eqn}
\text{LHS} &= \intinf \mathrm{d}t_1 \mathrm{d}t_2 \mathrm{d}t \mathrm{d}t' F_1(t_1) F_2(t_2) G_F(t_1, t, y_1) G_F(t', t_2, y_2) e^{- x_1 |t|} e^{- x_2 |t'|}  \\
&\qquad \times\big[  \sgn(t) + \sgn(t') \big] \int_{-i\infty}^{i\infty} \frac{\mathrm{d}w}{4\pi \mathrm{i}} e^{- w (|t| - |t'|)}
\end{eqn}
The $w$-integral gives $\frac12\delta(|t|-|t'|)$. Using the following relation 
\begin{eqn}\label{sgndelta1}
\big[ \sgn(t) + \sgn(t') \big] \delta(|t| - |t'|) 
&= 2 \sgn(t) \delta(t - t')
\end{eqn}
gives us the desired result,
\begin{eqn}
\text{LHS} &= \intinf \mathrm{d}t_1 \mathrm{d}t_2 \mathrm{d}t \mathrm{d}t' F_1(t_1) F_2(t_2) G_F(t_1, t, y_1) G_F(t', t_2, y_2) e^{- x_1 |t|} e^{- x_2 |t'|} \sgn(t)\delta(t - t') \\
&= \begin{tikzpicture}[baseline]
\draw (-1.5, 0) circle (0.5);
\draw (1.5, 0) circle (0.5);
\draw (-1, 0) -- (1,0);
\node at (0,0) {$\bullet$};
\node at (-1.5, 0) {$F_1$};
\node at (1.5, 0) {$F_2$};
\node at (-0.5, 0.25) {$y_1$};
\node at (0.5, 0.25) {$y_2$};
\node at (0, -0.25) {\scriptsize $x_1 + x_2$};
\end{tikzpicture}
\end{eqn}
hence completing the proof. Note that these fusion rules require the addition of two sets of graphs unlike those of those of the wavefunction for which a single set of graphs suffice \cite{Benincasa:2018ssx, Benincasa:2024leu, Chowdhury:2024snc}. 

We demonstrate this with an application to the bubble diagram,
\begin{eqn}\label{eq:bubblerec}
\int_{-i\infty}^{+i\infty} 
\frac{\mathrm{d}w}{4\pi \mathrm{i}} \ 
\begin{tikzpicture}[baseline]
\draw (0, -0.5) to[out = -180, in = 180] (-0.5, 0.5);
\draw (0, -0.5) to[out = 0, in = 0] (0.5, 0.5);
\node at (0, -0.5) {$\bullet$};
\node at (-0.5, 0.5) {$\circw$};
\node at (0.5, 0.5) {$\bullet$}; 
\node at (0, -0.75) {$x_2$};
\node at (-0.65, 0.75) {\scriptsize $x_1 + w$};
\node at (0.65, 0.75) {\scriptsize $x_3 - w$};
\end{tikzpicture}
+
\begin{tikzpicture}[baseline]
\draw (0, -0.5) to[out = -180, in = 180] (-0.5, 0.5);
\draw (0, -0.5) to[out = 0, in = 0] (0.5, 0.5);
\node at (0, -0.5) {$\bullet$};
\node at (-0.5, 0.5) {$\bullet$};
\node at (0.5, 0.5) {$\circw$}; 
\node at (0, -0.75) {$x_2$};
\node at (-0.65, 0.75) {\scriptsize $x_1 + w$};
\node at (0.65, 0.75) {\scriptsize $x_3 - w$};
\end{tikzpicture}
= \begin{tikzpicture}[baseline]
\draw (0,0) circle (0.5);
\node at (0, -0.5) {$\bullet$};
\node at (0, 0.5) {$\bullet$};
\node at (0, -0.75) {$x_2$};
\node at (0, 0.75) {$x_1 + x_3$};
\end{tikzpicture}.
\end{eqn}
The chain graphs in the LHS can be obtained from the recursions in equation \eqref{treerec1}. The integrals over $w$ can generically be evaluated using residues. We find that these recursion relations also expedite the computational time for evaluating loop integrands. 

By using a slightly different fusion rule we can also obtain a recursion relation for flat space correlators,
\begin{eqn}\label{eq:white_dot_fusion}
&\int_{-i\infty}^{+i\infty} \frac{\mathrm{d}w}{4\pi \mathrm{i}}  \quad 
\begin{tikzpicture}[baseline]
\draw (-1.5, 0) circle (0.5);
\node at (-1.5, 0) {$F_1$};

\draw (-1, 0) -- (0,0);
\node at (0, 0) {$\circw$};
\node at (0, -0.25) {\scriptsize $x_1+ w$};
\node at (-0.5, 0.25) {$y_1$};
\end{tikzpicture}
\begin{tikzpicture}[baseline]
\draw (1.5, 0) circle (0.5);
\node at (1.5, 0) {$F_2$};

\draw (1, 0) -- (0,0);
\node at (0, 0) {$\circw$};
\node at (0, -0.25) 
{\scriptsize $x_2 - w$};
\node at (0.5, 0.25) {$y_2$};
\end{tikzpicture}
+ 
\begin{tikzpicture}[baseline]
\draw (-1.5, 0) circle (0.5);
\node at (-1.5, 0) {$F_1$};
\node at (0, 0) {$\bullet$};
\draw (-1, 0) -- (0,0);
\node at (0, -0.25) {\scriptsize $x_1+ w$};
\node at (-0.5, 0.25) {$y_1$};
\end{tikzpicture}
\begin{tikzpicture}[baseline]
\draw (1.5, 0) circle (0.5);
\node at (1.5, 0) {$F_2$};
\node at (0, 0) {$\bullet$};
\draw (1, 0) -- (0,0);
\node at (0, -0.25) {\scriptsize $x_2 - w$};
\node at (0.5, 0.25) {$y_2$};
\end{tikzpicture}\\
&= 
\begin{tikzpicture}[baseline]
\draw (-1.5, 0) circle (0.5);
\draw (1.5, 0) circle (0.5);
\draw (-1, 0) -- (1,0);
\node at (0,0) {$\circw$};
\node at (-1.5, 0) {$F_1$};
\node at (1.5, 0) {$F_2$};
\node at (-0.5, 0.25) {$y_1$};
\node at (0.5, 0.25) {$y_2$};
\node at (0, -0.25) {\scriptsize $x_1 + x_2$};
\end{tikzpicture}
\end{eqn}
which follows by replacing the step \eqref{sgndelta1} with the following relation,
\begin{eqn}
\big[1 + \sgn(t) \sgn(t') \big] \delta(|t| - |t'|) = 2 \delta(t - t')
\end{eqn}
This establishes that the correlation functions of the field and the conjugate momentum satisfy a closed system of recursion relations relating the two.

\subsection*{Momentum-space Recursion}
The fusion rule also admit a momentum-space derivation. For example, for a black vertex with two edges in connection, we have
\begin{equation}
    \begin{tikzpicture}[baseline]
\draw (-1.5, 0) circle (0.5);
\draw (1.5, 0) circle (0.5);
\draw (-1, 0) -- (1,0);
\node at (0,0) {$\bullet$};
\node at (-1.5, 0) {$F_1$};
\node at (1.5, 0) {$F_2$};
\node at (-0.5, 0.25) {$y_1$};
\node at (0.5, 0.25) {$y_2$};
\node at (0, -0.25) {\scriptsize $x_1 + x_2$};
\end{tikzpicture}=\int_{-\infty}^\infty \frac{\mathrm{d}q_1}{\pi}\frac{\mathrm{d}q_2}{\pi}\frac{1}{q_1^2+y_1^2}\frac{1}{q_2^2+y_2^2}E_{x_1+x_2}(q_1+q_2)F_1(q_1)F_2(q_2),
\end{equation}
where $F_1(q_1)$ and $F_2(q_2)$ are the blobs connected to the black vertex with momentum $q_1$ and $q_2$ flowing into the vertex respectively. At the same time, 
\begin{equation}
    \begin{aligned}
        &\int_{-i\infty}^{+i\infty} \frac{\mathrm{d}w}{4\pi \mathrm{i}}  \quad 
\begin{tikzpicture}[baseline]
\draw (-1.5, 0) circle (0.5);
\node at (-1.5, 0) {$F_1$};
\draw (-1, 0) -- (0,0);
\node at (0, 0) {$\circ$};
\node at (0, -0.25) {\scriptsize $x_1+ w$};
\node at (-0.5, 0.25) {$y_1$};
\end{tikzpicture}
\begin{tikzpicture}[baseline]
\draw (1.5, 0) circle (0.5);
\node at (1.5, 0) {$F_2$};
\node at (0, 0) {$\bullet$};
\draw (1, 0) -- (0,0);
\node at (0, -0.25) {\scriptsize $x_2 - w$};
\node at (0.5, 0.25) {$y_2$};
\end{tikzpicture}
+ 
\begin{tikzpicture}[baseline]
\draw (-1.5, 0) circle (0.5);
\node at (-1.5, 0) {$F_1$};
\node at (0, 0) {$\bullet$};
\draw (-1, 0) -- (0,0);
\node at (0, -0.25) {\scriptsize $x_1+ w$};
\node at (-0.5, 0.25) {$y_1$};
\end{tikzpicture}
\begin{tikzpicture}[baseline]
\draw (1.5, 0) circle (0.5);
\node at (1.5, 0) {$F_2$};
\draw (1, 0) -- (0,0);
\node at (0, 0) {$\circ$};
\node at (0, -0.25) {\scriptsize $x_2 - w$};
\node at (0.5, 0.25) {$y_2$};
\end{tikzpicture}\\
=&\int_{-i\infty}^{+i\infty} \frac{\mathrm{d}w}{4\pi \mathrm{i}}\int_{-\infty}^\infty \frac{\mathrm{d}q_1}{\pi}\frac{\mathrm{d}q_2}{\pi}\frac{F_1(q_1)}{q_1^2+y_1^2}\frac{F_2(q_2)}{q_2^2+y_2^2}\left[H_{x_1+w}(q_1)E_{x_2-w}(q_2)+E_{x_1+w}(q_1)H_{x_2-w}(q_2)\right]\\
=&\int_{-\infty}^\infty \frac{\mathrm{d}q_1}{\pi}\frac{\mathrm{d}q_2}{\pi}\frac{1}{q_1^2+y_1^2}\frac{1}{q_2^2+y_2^2}E_{x_1+x_2}(q_1+q_2)F_1(q_1)F_2(q_2),
    \end{aligned}
\end{equation}
where in the last step we have utilized 
\begin{equation}
\int_{-i\infty}^{+i\infty} \frac{\mathrm{d}w}{4\pi \mathrm{i}}\left[H_{x_1+w}(q_1)E_{x_2-w}(q_2)+E_{x_1+w}(q_1)H_{x_2-w}(q_2)\right]=E_{x_1+x_2}(q_1+q_2).
\end{equation}
This shows that the vertex with two edges in connection can be decomposed into two vertices with one edge in connection, which is the same as the recursion  relation \eqref{eq:black_dot_fusion} for a black vertex with two edges in time space. The same procedure can be applied to a white vertex with two edges in connection \eqref{eq:white_dot_fusion}, by using the following integral identity:
\begin{equation}
    \int_{-i\infty}^{+i\infty} \frac{\mathrm{d}w}{4\pi \mathrm{i}}\left[H_{x_1+w}(q_1)H_{x_2-w}(q_2)+E_{x_1+w}(q_1)E_{x_2-w}(q_2)\right]=H_{x_1+x_2}(q_1+q_2).
\end{equation}
Following the similar procedure as the $n$-chain case, we can also derive the momentum-space representation for any polygon graph recursively. We demonstrate this with an example. For $G$ being an $n$-site polygon the polynomials $\mathcal P_u(q), \mathcal P_v(q)$ defined in \eqref{eq:GBWgeneral1} are given as,
\begin{equation}
    \mathcal{P}_v(q)=q_v-q_{v-1},\quad\text{with}\quad q_0=q_n,
\end{equation}thus with all vertices being black, we have the following integral representation in momentum-space,
\begin{equation}\label{eq:npolygonblack}
    \langle\text{$n$-gon}\rangle_{n;0}=\int_{-\infty}^\infty \prod_{i=1}^n\frac{-2\mathrm{i}(q_i-q_{i-1})}{(q_i-q_{i-1})^2+x_i^2}\prod_{e=1}^n\frac{\mathrm{d}q_e}{\pi}\frac{1}{q_e^2+y_e^2},
\end{equation}
and similarly for all vertices being white,
\begin{equation}
    \langle\text{$n$-gon}\rangle_{0;n}=\int_{-\infty}^\infty \prod_{j=1}^n\frac{2x_j}{(q_j-q_{j-1})^2+x_j^2}\prod_{e=1}^n\frac{\mathrm{d}q_e}{\pi}\frac{1}{q_e^2+y_e^2},
\end{equation}
Via the fusion relation we can derive the result of these integrals using the tree level graphs. For example, the bubble diagram with black vertices which satisfies the recursion shown in \eqref{eq:bubblerec} evaluates to,
\begin{equation}\label{eq:buble_integrand}
    \left< \text{$2$-gon}\right> _{2;0}=\frac{4(y_1+y_2)}
{y_1y_2(x_1+x_2)(x_1+y_1+y_2)(x_2+y_1+y_2)}.
\end{equation}
Obtaining such closed form expressions is one of the main practical advantages of the direct representation introduced in section \ref{sec:directrep}.

\subsection{Melonic Diagrams}\label{sssec:melon}
As emphasized in \cite{Donath:2024utn, Arkani-Hamed:2025mce}, melonic diagrams for field correlators collapse to a simple dressed exchange graph,
\begin{eqn}
\begin{tikzpicture}[baseline]
\draw (0,0) circle (0.5);
\node at (-0.5, 0) {$\circw$};
\node at (0.5, 0) {$\circw$};
\node at (-0.75, 0) {$x_1$};
\node at (0.8, 0) {$x_2$};
\node at (0, 0.75) {$y_1$};
\node at (0,0.1) {$\vdots$};
\node at (0, -0.75) {$y_n$};
\end{tikzpicture}
= \frac{\Sigma_iy_i}{\prod_i y_i} 
\begin{tikzpicture}[baseline]
\draw (-0.5, 0) -- (0.5, 0);
\node at (-0.5, 0) {$\circw$};
\node at (0.5, 0) {$\circw$};
\node at (-0.75, 0) {$x_1$};
\node at (0.8, 0) {$x_2$};
\node at (0, 0.25) {$\Sigma_i y_i$};

\end{tikzpicture}~.
\end{eqn}
This relation is derived by using the following property of the Feynman propagator,
\begin{eqn}
G_F(t_1, t_2; y_1) G_F(t_1, t_2; y_2) = \frac{y_1 + y_2}{y_1 y_2} G_F(t_1, t_2; y_1 + y_2)
\end{eqn}
Since this formula only relies on the properties of the Feynman propagator they are also true for Melonic diagrams for correlators of the conjugate momentum,
\begin{eqn}
\label{eq:dS_melonic}
\begin{tikzpicture}[baseline]
\node at (-0.5, 0) {$\bullet$};
\node at (0.5, 0) {$\bullet$};
\draw (0,0) circle (0.5);
\node at (-0.75, 0) {$x_1$};
\node at (0.8, 0) {$x_2$};
\node at (0, 0.75) {$y_1$};
\node at (0,0.1) {$\vdots$};
\node at (0, -0.75) {$y_n$};
\end{tikzpicture}
= \frac{\Sigma_iy_i}{\prod_i y_i} 
\begin{tikzpicture}[baseline]
\node at (-0.5, 0) {$\bullet$};
\node at (0.5, 0) {$\bullet$};
\draw (-0.5, 0) -- (0.5, 0);
\node at (-0.75, 0) {$x_1$};
\node at (0.8, 0) {$x_2$};
\node at (0, 0.25) {$\Sigma_i y_i$};

\end{tikzpicture}
\end{eqn}

\subsection{Energy Expansions}
We now turn to the singularity structure of the correlators.
The observables considered here have simple poles whenever the sum of energies entering a vertex, or a connected collection of vertices, vanishes; see for example \eqref{chainexample1}.
These are the familiar partial-energy singularities (for instance $x_1+y$ in \eqref{eq:2chain}).
A particularly important special case is the total-energy pole, where the sum of all external energies vanishes, whose residue reproduces the flat space amplitude \cite{Raju:2012zr}.
Correlators of fields are known to factorize on these poles \cite{Arkani-Hamed:2017fdk, Benincasa:2022gtd, Benincasa:2024leu, Chowdhury:2024snc}.
Here we study the corresponding series expansion around such singularities.
This analysis relies crucially on the time integral representation \eqref{eq:time_representation}, since the pole structure is not manifest in the dressing-rule representation\footnote{While the flat space limit can be derived directly from the dressing rules \cite{Chowdhury:2026upp}, doing so requires taking discontinuities with respect to external energies, analogous to an LSZ-type extraction. The singularity structure itself is much more transparent in time space.}.

To be concrete we shall consider the expansion about the total energy pole and the discussion for the expansions about other partial energy singularities follows in a similar manner. Consider the expression for the general correlator \eqref{eq:time_representation} for tree level graphs (the loop level generalization is straightforward and discussed below),
\begin{eqn}
\braket{G}_{B;W}=\intinf \mathrm{d}t_1 e^{ - x_1 |t_1|} \sgn(t_1) \intinf \prod_{i=2}^B \mathrm{d}s_i \sgn(t_1 + s_i)e^{-x_i|t_1 + s_i|} \prod_{u=B+1}^{W} \mathrm{d}s_j e^{-x_j |t_1 + s_j|} \prod_{e\in E} G_F(|s_{e}|; y_e),
\end{eqn}
where we have used $t_i = t_1 + s_i \ \forall \ i \geq 2$. We have singled out $t_1$ by making the choice that it is the black vertex. However this is completely unnecessary as we reach the same conclusion by making any other choice. It is convenient to first perform the following change of variables \cite{Arkani-Hamed:2025mce},
\begin{eqn}
t_1 = \frac{1}{\e}\tau , \qquad x_i = \e \bar x_i
\end{eqn}
This effectively zooms into the limit with all $x_i \to 0$ and makes the $\e$-dependence manifest, 
\begin{eqn}\label{eq:Gblackeps}
\braket{G}_{B;W} = \intinf \frac{\mathrm{d}\tau}{\e} e^{ - \bar x_1 |\tau|} \sgn(\tau) \intinf \prod_{i = 2}^B \mathrm{d}s_i \sgn(\tau + \e s_i) e^{- \bar x_i |\tau +  \e s_i|} \prod_{u=B+1}^{W} \mathrm{d}s_j e^{- \bar x_j |\tau +  \e s_j|}  \prod_{e\in E} G_F(|s_{e}|; y_e)
\end{eqn}
The expansion in $\e$ can be now performed in a straightforward manner by using,
\begin{align}
e^{- \bar x |\tau + \e s|} &= e^{- \bar x |\tau|} \Big[ 1 - \bar x s \e \sgn(\tau) + \frac12 \bar x s^2 \e^2 (\bar x - 2 \delta(\tau)) + O(\e^3)\Big], \\
\sgn(\tau + \e s) e^{- \bar x |\tau + \e s|} &= \sgn(\tau) e^{- \bar x|\tau|} + \e s \big[ - \bar x e^{- \bar x|\tau|} + 2 \delta(\tau) \big] + \frac{1}{2} \e^2 s^2 \big[ \bar x^2 \sgn(\tau) e^{- \bar x |\tau|} + 2 \delta'(\tau) \big] + O(\e^3)\nno
\end{align}
The leading order term in $\e$ of \eqref{eq:Gblackeps} recovers the usual total energy pole,
\begin{eqn}\label{eq:expansionleading1}
\braket{G}_{B; W}^{(-1)} &= \intinf  \frac{\mathrm{d}\tau}{\e} e^{- (\bar x_1 + \cdots + \bar x_{B+W})|\tau| } \sgn^B(\tau) \intinf \prod_{i  = 2}^{B+W} \mathrm{d}s_i  \prod_{e \in E} G_F(|s_e|; y_e) \\
&= \frac{2}{x_1 + \cdots + x_{B+W}} \prod_{e \in E} \frac{2}{y_e^2}
\end{eqn}
where we have assumed $B$ is even, otherwise the graph vanishes (see section \ref{ssec:oddchain}). The leading order term is independent of the type of vertices which is expected as it should reproduce the amplitude in the $x_i \to 0$ limit.

The subleading order term in $\e$ in \eqref{eq:Gblackeps} vanishes. To see this explicitly we expand the answer to $O(\e)$ and obtain,
\begin{align}
\braket{G}_{B; W}^{(0)} &= \intinf \mathrm{d}\tau e^{- (\bar x_1 + \cdots + \bar x_{B-1})|\tau|}   \Bigg\{ \sgn^{B+1}(\tau) s_W x_{W} \int \prod_{i = 2}^{B+W-1} \mathrm{d}s_i \\
&\qquad+ \sgn^{B-1}(\tau) \big[ - \bar x_B e^{- \bar x_B |\tau|} + 2 \delta(\tau) \big]  \intinf \prod_{i = 2}^{B+W} \mathrm{d}s_i s_B \Bigg\} \prod_{e \in E} G_F(|s_e|; y_e) + \text{perms}\nno
\end{align}
where the permutations are taken over all possible terms that contribute at $O(\e)$.
Due to the odd number of $\sgn(\tau)$ insertions, the $\tau$-integrand is odd and the integral vanishes.
Therefore the subleading term in the energy expansion vanishes for arbitrary graphs built from black and white vertices.
The same conclusion also holds for loop integrands.
At the level of the final  answer, the only difference appears in the leading term \eqref{eq:expansionleading1}: instead of the factor $\prod_e y_e^{-2}$ one obtains the flat space amplitude in the $x_i\to 0$ limit after integrating over the loop energy $l_0$.
The vanishing of the subleading term for correlators with all white vertices was already noted in \cite{Arkani-Hamed:2025mce}.
An analogous statement holds for expansions about partial-energy singularities, with the only change that the leading term factorizes into a lower-point correlator times an amplitude.

Higher orders in the expansion are more subtle. Already at $O(\e^2)$ one encounters products of delta functions such as $\delta^2(\tau)$, signaling non-analytic behavior in the small-\(\e\) expansion.
For graphs with only white vertices this phenomenon first appears at $O(\e^4)$, whereas in the mixed black/white case it already arises at $O(\e^2)$.
Even so, the leading and subleading terms are controlled universally.

The results of this section exhibit a common pattern. The black/white representation turns operations that are obscure in the original in-in expression into elementary graph moves wheref odd sectors vanish by parity, trees reduce by removing leaves, more general graphs can be glued by fusion, melons collapse to effective edges, and the first terms in energy-singularity expansions are universal. We next ask whether this integrand-level economy survives the remaining energy integrations.


\section{Towards Simplicity After Integration: Alphabets and Symbols}\label{sec:symbol_simplicity}

In the following section we move beyond the pre-integrated representation and ask how much of the simplicity discussed above survives after the energy integrals are performed.
At this stage the question becomes sharper: if the correlator is intrinsically simpler, its symbol should not merely reproduce the wavefunction alphabet after cancellations, but should be governed by a smaller or differently organized set of letters.
Our aim is not yet a complete classification, but rather to exhibit concrete examples showing that this expectation is borne out. 

Throughout this section, we use the notation $\langle G\rangle_{\rm dS}$ to denote the highest weight part, say, parts corresponding to the maximal number of black dots configuration, in the correlation function of a graph $G$. The lower weight sectors can be computed by the recursion relation \eqref{eq:recusion_formula} and thus do not contribute extra letters compared with the highest weight parts. We begin with a few case studies that illustrate how the energy integrals are carried out.
For example, from \eqref{chain_recursion} the in-in correlator of the 2-site chain is:
\begin{eqn}
    \langle\text{2-chain}\rangle_{\rm dS}&=\int_0^\infty \mathrm{d}s_1\mathrm{d}s_2\begin{tikzpicture}[baseline]
\node at (-.5, 0) {$\bullet$};
\node at (.5, 0) {$\bullet$};
\draw (-.5, 0) -- (.5, 0);
\node at (-.5, -0.25) {$x_1$};
\node at (.5, -0.25) {$x_2$};
\node at (0, 0.25) {$y$};
\end{tikzpicture}\bigg|_{\substack{x_1\to x_1+s_1\\
x_2\to x_2+s_2}}\\&=\int_0^\infty \mathrm{d}s_1\mathrm{d}s_2\frac{4}{(x_1+x_2+s_1+s_2)(x_1+s_1+y)(x_2+s_2+y)}.
\end{eqn}
By partial-fractioning the denominators, this integral can be written as
\begin{eqn}
    \langle\text{2-chain}\rangle_{\rm dS}&=\frac{2}{y}\int_0^\infty \mathrm{d}s_2\left(\frac{1}{x_2+s_2-y}-\frac{1}{x_2+s_2+y}\right)\int_0^\infty \mathrm{d}s_1\left(\frac{1}{x_1+s_1+y}-\frac{1}{x_1+x_2+s_1+s_2}\right)\\
    &=\frac{2}{y}\int_0^\infty \mathrm d_{s_2}\log\left(\frac{x_2+s_2-y}{x_2+s_2+y}\right)\mathrm d_{s_1}\log\left(\frac{x_1+s_1+y}{x_1+x_2+s_1+s_2}\right),
\end{eqn}
where $\mathrm d_{s_i}$ denotes taking differential only in accordance with the variable $s_i$.
We use the method proposed in \cite{Caron-Huot:2011dec,He:2020uxy} and automatic symbol recursion codes \texttt{abcode.m} and \texttt{NumPolyLog.m} in \cite{doubilee_files} to obtain the symbolic results:
\begin{equation}
    \langle\text{2-chain}\rangle_{\rm dS}=\frac{2}{y}\left(\frac{x_2+y}{x_1+x_2}\right)\otimes\left(\frac{x_1-y}{x_1+y}\right)+\frac{2}{y}\left(\frac{x_1+y}{x_1+x_2}\right)\otimes\left(\frac{x_2-y}{x_2+y}\right).
\end{equation}
By collecting all the symbol letters, we find that the alphabet of the in-in correlator is the same as that of the 2-site chain wavefunction {\cite{Arkani-Hamed:2017fdk,Hillman:2019wgh,Arkani-Hamed:2023decorr,Arkani-Hamed:2023kflow,Baumann:2024loopflow,Mazloumi:2025pmx,Capuano:2025myy,Paranjape:2026htn,Capuano:2026pgq}}.  
The same procedure also applies at loop level. For the bubble diagram, whose integrand was given in \eqref{eq:buble_integrand}, the $\mathrm{d}\log$ representation reads
\begin{eqn}
    \langle\text{2-gon}\rangle_{\rm dS}&=\int\mathrm{d}^3l\int_{0}^\infty\mathrm{d}s_1\mathrm{d}s_2\begin{tikzpicture}[baseline]
\draw (0,0) circle (0.5);
\node at (-.5, 0) {$\bullet$};
\node at (0.5, 0) {$\bullet$};
\node[right] at (0.5, 0) {$x_2$};
\node[left] at (-0.5,0 ) {$x_1 $};
\node[above] at (0, 0.5) {$y_1$};
\node[below] at (0,-.5 ) {$y_2 $};
\end{tikzpicture}\Bigg|_{\substack{x_1\to x_1+s_1\\ x_2\to x_2+s_2}}\\
&=\frac{2}{y_1y_2}\int\mathrm{d}^3l\int_0^\infty \mathrm d_{s_2}\log\left(\frac{x_2+s_2-y_1-y_2}{x_2+s_2+y_1+y_2}\right)\mathrm{d}_{s_1}\log\left(\frac{x_1+s_1+y_1+y_2}{x_1+x_2+s_1+s_2}\right),
\end{eqn}
where $y_1=|\vec l|$ and $y_2=|\vec l+\vec k|$, with $\vec k$ the external momentum flowing into the loop. The spatial part of the loop momentum is left unintegrated, since integrating it out produces an ultraviolet divergence that requires renormalization; see Appendix \ref{app:loop} for further discussion. Performing the $s$-integrals at symbol level, we obtain
\begin{eqn}
    \langle\text{2-gon}\rangle_{\rm dS}=\frac{2}{y_1y_2}\int\mathrm{d}^3l\left[\left(\frac{x_2+y_1+y_2}{x_1+x_2}\right)\otimes\left(\frac{x_1-y_1-y_2}{x_1+y_1+y_2}\right)+\left(\frac{x_1+y_1+y_2}{x_1+x_2}\right)\otimes\left(\frac{x_2-y_1-y_2}{x_2+y_1+y_2}\right)\right].
\end{eqn}
We see that 
\begin{equation}
    \langle\text{2-gon}\rangle_{\rm dS}=\frac{y_1+y_2}{y_1y_2}\langle\text{2-chain}\rangle_{\rm dS}\Big|_{y\to y_1+y_2},
\end{equation}
which agrees with \eqref{eq:dS_melonic}. In contrast with the wavefunction alphabet \cite{Baumann:2024loopflow,Paranjape:2026htn}, we see that some letters are already absent for even-point graphs at loop level, as depicted below together with their corresponding tubings\footnote{Readers unfamiliar with the formulation of symbol letters in terms of tubings may consult \cite{Arkani-Hamed:2023decorr,Arkani-Hamed:2023kflow} for a pedagogical introduction.}:
\begin{eqn}\label{eq:bubble_letter_lost}
    \raisebox{-.5cm}{\includegraphics[scale=.55]{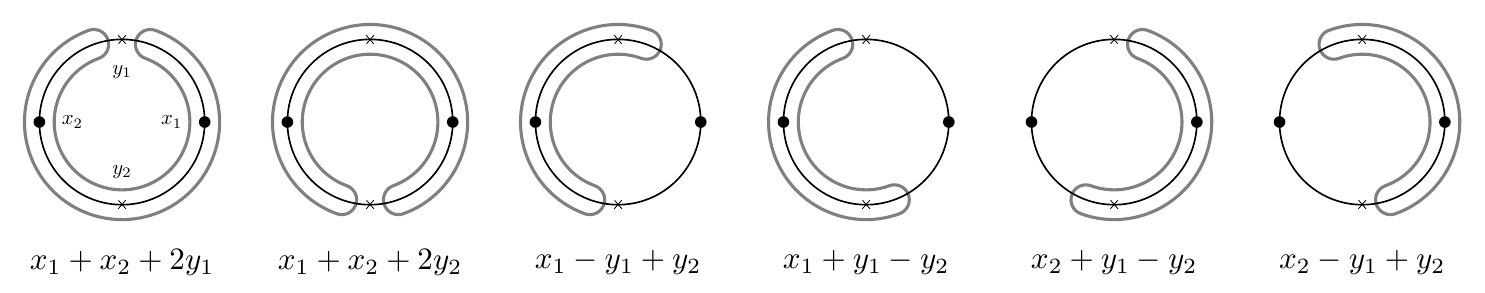}}
\end{eqn}
The odd-point case is more interesting. Since not all $n$ vertices can be taken to be black, otherwise the correlator vanishes by Section \ref{ssec:oddchain}, there are at most $(n-1)$ energy integrals, leading to a drop in the maximal transcendental weight for graphs with an odd number of vertices. The first non-trivial example is the 3-site chain. One contribution is obtained by taking $x_1$ and $x_2$ to be connected by dashed lines and $x_3$ by a dotted line:
\begin{eqn}
    \langle\vcenter{\hbox{\begin{tikzpicture}
\fill[]  (-0.4,0) circle  (2pt);
\node at (0, 0) {$\bullet$};
\filldraw[fill=gray!30,thick]  (0.4,0) circle  (2pt);
\draw (-.4, 0) -- (.34, 0);
\end{tikzpicture}}}\rangle_{\rm dS}&=\int_0^\infty\mathrm{d}s_1\mathrm{d}s_2
\begin{tikzpicture}[baseline]
\node at (-1, 0) {$\bullet$};
\node at (0, 0) {$\bullet$};
\filldraw[fill=gray!30,thick]  (1,0) circle  (2pt);
\draw (-1, 0) -- (.94, 0);
\node[below] at (-1,0) {$x_1$};
\node[below] at (0,0) {$x_2$};
\node[below] at (1,0) {$x_3$};
\node[above] at (-.5,0) {$y_1$};
\node[above] at (.5,0) {$y_2$};
\end{tikzpicture}\bigg|_{\substack{x_1\to x_1+s_1\\ x_2\to x_2+s_2}}\\
&=\frac{-2\pi}{y_1y_2}\left[\left(\frac{x_2+y_1+y_2}{x_1+x_2+y_2}\right)\otimes\left(\frac{x_1-y_1}{x_1+y_1}\right)+\left(\frac{x_1+y_1}{x_1+x_2+y_2}\right)\otimes\left(\frac{x_2-y_1+y_2}{x_2+y_1+y_2}\right)\right],
\end{eqn}
Here the gray vertex $\vcenter{\hbox{\begin{tikzpicture}[baseline]
\filldraw[fill=gray!30,thick]  (1,0) circle  (2pt);
\end{tikzpicture}}}$ denotes a point linked by a dotted line. The second contribution is obtained by exchanging $x_1\leftrightarrow x_3$ and $y_1\leftrightarrow y_2$:
\begin{eqn}
   \langle\vcenter{\hbox{\begin{tikzpicture}
\fill[]  (0.4,0) circle  (2pt);\node at (0, 0) {$\bullet$};
\filldraw[fill=gray!30,thick]  (-0.4,0) circle  (2pt);
\draw (-.34, 0) -- (.4, 0);
\end{tikzpicture}}}\rangle_{\rm dS} =\langle\vcenter{\hbox{\begin{tikzpicture}
\fill[]  (-0.4,0) circle  (2pt);\node at (0, 0) {$\bullet$};
\filldraw[fill=gray!30,thick]  (0.4,0) circle  (2pt);
\draw (-.4, 0) -- (.34, 0);
\end{tikzpicture}}}\rangle_{\rm dS}\bigg|_{\substack{x_1\leftrightarrow x_3\\ y_1\leftrightarrow y_2}}.
\end{eqn}
The 3-site chain correlator is then given by the sum of these two contributions, up to the trivial all-gray part,
\begin{equation}
    \langle\text{3-chain}\rangle_{\rm dS}= \langle\vcenter{\hbox{\begin{tikzpicture}
\fill[]  (0.4,0) circle  (2pt);\node at (0, 0) {$\bullet$};
\filldraw[fill=gray!30,thick]  (-0.4,0) circle  (2pt);
\draw (-.34, 0) -- (.4, 0);
\end{tikzpicture}}}\rangle_{\rm dS} +\langle\vcenter{\hbox{\begin{tikzpicture}
\fill[]  (-0.4,0) circle  (2pt);\node at (0, 0) {$\bullet$};
\filldraw[fill=gray!30,thick]  (0.4,0) circle  (2pt);
\draw (-.4, 0) -- (.34, 0);
\end{tikzpicture}}}\rangle_{\rm dS},
\end{equation}
whose alphabet is observed to miss the following 4 letters from the wavefunction side:
\begin{eqn}\label{eq:three_chain_letter_lost}
    \vcenter{\hbox{\begin{tikzpicture}
        \begin{scope}
            \node at (-1, 0) {$\bullet$};
\node at (0, 0) {$\bullet$};
\node at (1, 0) {$\bullet$};
\draw (-1, 0) -- (1, 0);
\node[] at (-.5,0) {$\times$};
\node[] at (.5,0) {$\times$};
\draw[gray,line width=1.5pt,rounded corners=6pt]
(-.7,-0.2) rectangle (.7,.2);
\node at (0,-.5) {$x_2-y_1-y_2$};
\node[above] at (-1,0.2) {$x_1$};
\node[above] at (-.5,0.2) {$y_1$};
\node[above] at (0,0.2) {$x_2$};
\node[above] at (.5,0.2) {$y_2$};
\node[above] at (1,0.2) {$x_3$};
        \end{scope}
        \begin{scope}[xshift=3cm]
            \node at (-1, 0) {$\bullet$};
\node at (0, 0) {$\bullet$};
\node at (1, 0) {$\bullet$};
\draw (-1, 0) -- (1, 0);
\node[] at (-.5,0) {$\times$};
\node[] at (.5,0) {$\times$};
\draw[gray,line width=1.5pt,rounded corners=6pt]
(-1.2,-0.2) rectangle (.7,.2);
\node at (0,-.5) {$x_1+x_2-y_2$};
        \end{scope}
        \begin{scope}[xshift=6cm]
            \node at (-1, 0) {$\bullet$};
\node at (0, 0) {$\bullet$};
\node at (1, 0) {$\bullet$};
\draw (-1, 0) -- (1, 0);
\node[] at (-.5,0) {$\times$};
\node[] at (.5,0) {$\times$};
\draw[gray,line width=1.5pt,rounded corners=6pt]
(-.7,-0.2) rectangle (1.2,.2);
\node at (0,-.5) {$x_2+x_3-y_1$};
        \end{scope}
        \begin{scope}[xshift=9cm]
            \node at (-1, 0) {$\bullet$};
\node at (0, 0) {$\bullet$};
\node at (1, 0) {$\bullet$};
\draw (-1, 0) -- (1, 0);
\node[] at (-.5,0) {$\times$};
\node[] at (.5,0) {$\times$};
\draw[gray,line width=1.5pt,rounded corners=6pt]
(-1.2,-0.2) rectangle (1.2,.2);
\node at (0,-.5) {$x_1+x_2+x_3$};
        \end{scope}
    \end{tikzpicture}}}.
\end{eqn}
In this way, we can compute multiple polylogarithmic functions and their symbols directly from the $\mathrm{d}\log$ representations, both for wavefunction coefficients and for de Sitter correlators. In practice the computation becomes more challenging as the weight increases, but we are still able to obtain explicit symbols, for example for the $n$-chain and $n$-gon up to $n=5$\footnote{In addition to the examples discussed in the main text,
other symbol results, including the 4-site chain, 4-site star, 5-site star, box diagram and pentagon diagram, can be found in the ancillary files.}.

Our goal in this section is not a complete analysis of these symbols, but rather to present evidence that the symbol alphabet of the de Sitter correlator can be simpler than that of the corresponding wavefunction.
\subsection{Tree Diagrams}
For the highest-weight part of a tree level correlator, no letters are missing for even-point trees, since all vertices can be connected by dashed lines and no gray vertices appear in this configuration. The degeneration of the alphabet therefore begins only for odd-point trees.

As shown in \eqref{eq:three_chain_letter_lost}, the 3-chain has four missing letters, and in tubing language each of them encloses two marked points. We now extend this discussion to a slightly more complicated example, the five-star diagram. There are three distinct choices for the gray vertex, denoted by $t_1$, $t_2$, and $t_3$, and the weight-four part of the correlator is obtained by summing over these three contributions,
\begin{eqn}
    \vcenter{\hbox{\includegraphics[scale=.75]{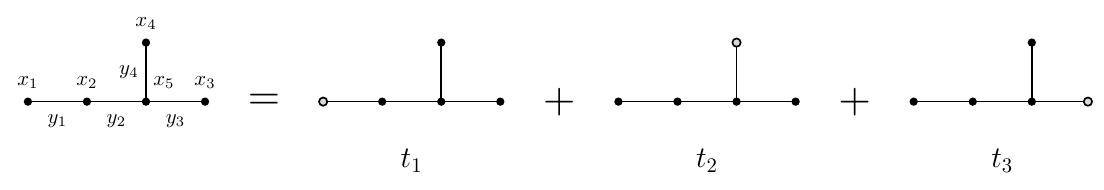}}}.
\end{eqn}
We find that, taken together, these contributions miss eight letters relative to the five-star wavefunction, as summarized below:
\begin{eqn}
    \vcenter{\hbox{\includegraphics[scale=.7]{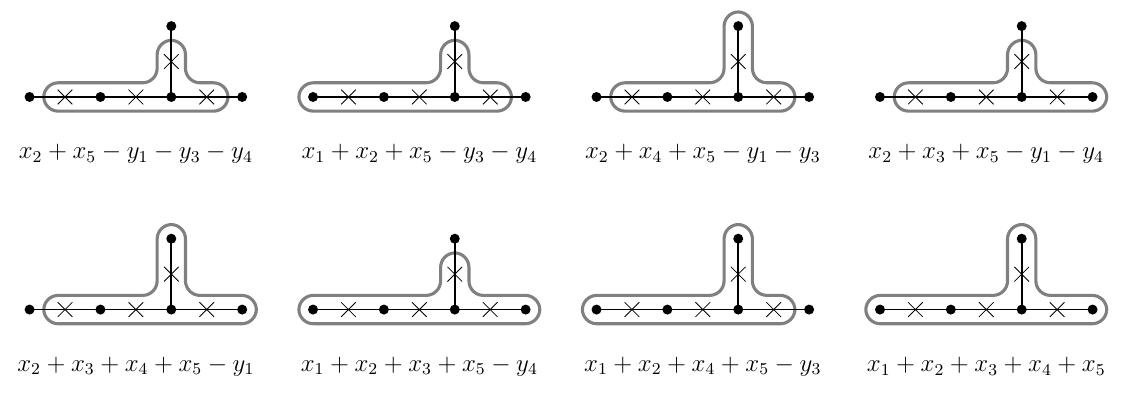}}}
\end{eqn}
It is then clear that the missing letters are precisely those that include all marked points on $y_1$, $y_2$, $y_3$, and $y_4$.

This pattern generalizes directly to any $n$-tree diagram with $\ell$ leaves. For each leaf vertex $x_i$ attached to the edge $y_i$, there is a unique arrangement $t_i$ in which $x_i$ is gray and all other vertices are black. In each such arrangement, the neighboring vertex $x_j$ is shifted from $x_j$ to $x_j+y_i$ according to \eqref{eq:recusion_formula}. Therefore, no matter how one chooses tubings in $t_i$, they can enclose at most $(n-2)$ marked points, namely all marked points except the one on $y_i$. This shows that, in contrast with the odd-point tree wavefunction, the correlator loses precisely those letters that contain all marked points attached to leaves. For each leaf, the corresponding missing letter can be written in two equivalent ways: either by including the leaf energy itself or by replacing it with the negative edge variable, thus we conclude: For any odd-n-tree graph 
which has $\ell$ leaves $x_1,\, x_2,\,\dots,\,x_\ell$ connected by edges $y_1,\,y_2,\,\dots,\, y_\ell$, there are $2^\ell$ missing letters of this correlator compared with the wavefunction, which are \begin{equation}
        \sum_{i=\ell+1}^nx_i+\sum_{j=1}^\ell\begin{cases}
            x_j,&\text{or}\\
            (-y_j).
        \end{cases}
    \end{equation}

\subsection{Polygonal Diagrams}
At loop level the situation becomes richer. For polygon diagrams with $n$ vertices, the structure naturally splits according to the parity of $n$. Odd-point and even-point polygons require slightly different representations, reflecting the alternating pattern of dashed and dotted sectors around the loop. It is therefore useful to discuss the two cases separately, even though they are governed by the same underlying gluing logic.

For an odd-point polygon, not all vertices can be connected by dashed lines, so there must be at least one dotted line, or equivalently one gray vertex. This leads to the decomposition of the correlator described in \eqref{eq:recusion_formula}. The contributions containing the maximal number of vertices, namely $(n-1)$, are precisely the arrangements that already appeared in the tree level discussion. The highest-weight part of the correlator is therefore obtained by summing over these configurations,
\begin{eqn}
    \vcenter{\hbox{\includegraphics[scale=.8]{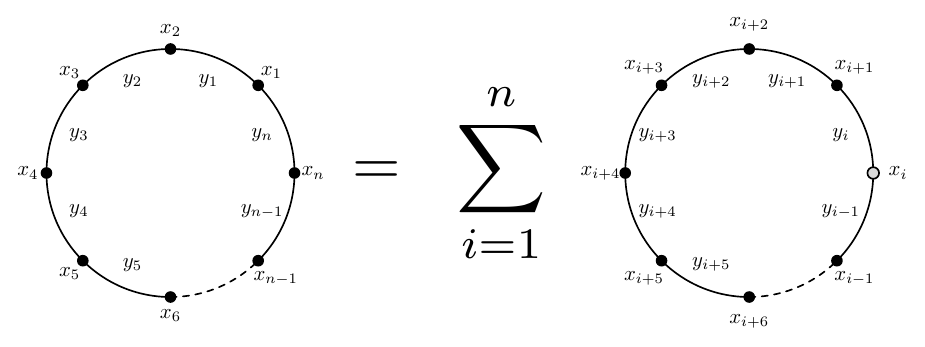}}}.
\end{eqn}
The simplest example is the triangle. Applying \eqref{eq:recusion_formula}, we find that its integrand differs from that of the 2-site chain only by the shifts $x_1\to x_1+y_3$ and $x_2\to x_2+y_2$. The symbol of the triangle can therefore be read off directly by these replacements together with cyclic permutations:
\begin{eqn}
    \langle\text{3-gon}\rangle_{\rm dS}=\frac{-\pi}{y_2y_3}\langle\text{2-chain}\rangle_{\rm dS}\Big|_{\substack{x_1\to x_1+y_3\\ x_2\to x_2+y_2}}+\text{cyclic}(1,2,3).
\end{eqn}
Comparing with the alphabet of the triangle wavefunction, we can classify the missing letters according to the number of vertices and marked points they contain, as summarized in table \ref{table:oddgon_letter}.

\begin{table}
\centering
    \begin{tblr}
{c|[dotted]c|[dotted]c}
    \hline[1pt]
    \diagbox[]{\# of ``\textbullet''}{\# of ``$\times$''}&2&3\\
    \hline[] 1&$\vcenter{\hbox{\includegraphics[scale=.8]{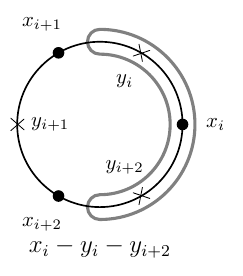}}}$&---\\\hline[dashed]2&$\vcenter{\hbox{\includegraphics[scale=.8]{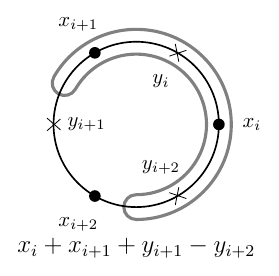}}}$$\vcenter{\hbox{\includegraphics[scale=.8]{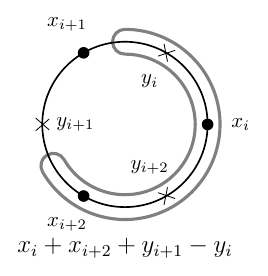}}}$&$\vcenter{\hbox{\includegraphics[scale=.8]{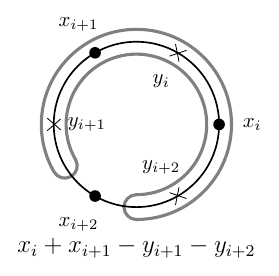}}}$\\\hline[dashed]
    3&$\vcenter{\hbox{\includegraphics[scale=.8]{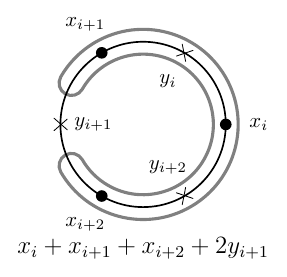}}}$&$\vcenter{\hbox{\includegraphics[scale=.8]{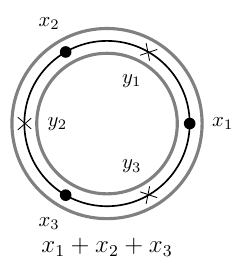}}}$\\\hline[]\SetCell[c=3]{m} Total number of missing letters\ :\qquad $1\times3+0+2\times3+1\times3+1\times3+1=16$&&\\\hline[1pt]
\end{tblr}
\caption{Missing letters of the 3-gon sorted by the number of internal vertices and marked points.}
\label{table:oddgon_letter}
\end{table}
The same logic extends directly to odd $n$-gons. If $x_i$ is taken to be gray, then, in accordance with \eqref{eq:recusion_formula}, the adjacent vertices are shifted as $x_{i+1}\to x_{i+1}+y_{i}$ and $x_{i-1}\to x_{i-1}+y_{i-1}$, with cyclic indices understood. As a result, any letter appearing in the final answer can contain at most $(n-2)$ marked points. Thus the odd-point polygon correlator loses precisely those letters that contain either $n$ or $(n-1)$ marked points. Summing over all such letters, we find that there are $5n+1$ missing letters for the odd-point polygon correlator relative to the wavefunction. They can be written as
\begin{equation}
  \begin{matrix}
      \displaystyle\sum_{i=1}^nx_i,&\displaystyle\sum_{i=j+1}^{j-1}x_i-y_j-y_{j-1},&\displaystyle\sum_{i=1}^nx_i+2y_j,\\\displaystyle\sum_{i=j+1}^{j-1}x_i+y_j-y_{j-1},&\displaystyle\sum_{i=j+1}^{j-1}x_i+y_j-y_{j+1},&\displaystyle\sum_{i=j+2}^{j-1}x_i-y_{j-1}-y_{j+1},
  \end{matrix}  
\end{equation}
\begin{table}
   \begin{tblr}
   {c|[dotted]c|[dotted]c}
    \hline[1pt]
    \diagbox[]{\# of ``\textbullet''}{\# of ``$\times$''}&2&\qquad3\\
    \hline[] 1&---&\qquad---\qquad\\\hline[dashed]2&$\vcenter{\hbox{\includegraphics[scale=.8]{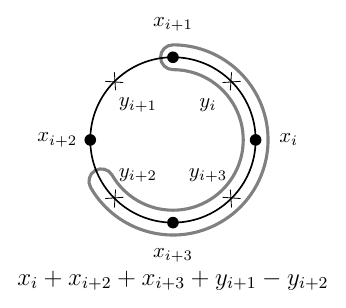}}}$$\vcenter{\hbox{\includegraphics[scale=.8]{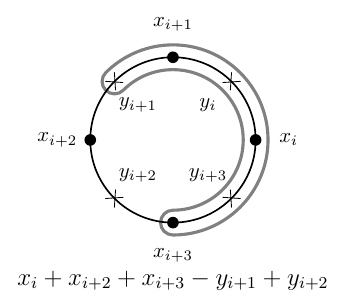}}}$&\qquad---\qquad\\\hline[dashed]
    3&$\vcenter{\hbox{\includegraphics[scale=.8]{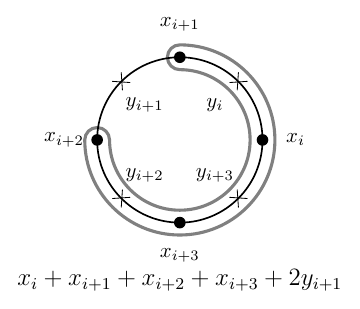}}}$&\qquad---\qquad\\\hline[]\SetCell[c=3]{m} Total number of missing letters\ :\qquad $1\times4+2\times4=12$&&\\\hline[1pt]
\end{tblr}
\caption{Missing letters of the 4-gon by the number of internal vertices and marked points.}
\label{table:evengon_letters}
\end{table}
where all indices are understood cyclically, and sums such as $\sum_{i=j+1}^{j-1}$ are taken along the oriented polygon.

For even-point polygons the story changes in an essential way: unlike the odd case, one can draw a fully dashed configuration with no obligatory gray vertex. As a result, the decomposition into sectors is less rigid, and the highest-weight part is no longer captured solely by the tree-like configurations inherited from the odd-point analysis. The integrated alphabet is therefore less reduced, with the correlator receiving contributions both from the fully dashed sector and from mixed sectors containing gray vertices. In this subsection we only sketch this structure and record the main qualitative difference from the odd-point case; a complete analysis of the even-point polygon alphabet will be left for future work.

This behavior is already visible in the simplest example. As noted in \eqref{eq:bubble_letter_lost}, the bubble diagram already exhibits missing letters relative to the 2-site chain. These letters can be grouped into three classes,
\begin{eqn}
    \sum_{j=1}^nx_j+2y_i,\quad \sum_{j=i+1}^{i-1}x_j+y_i-y_{i-1},\quad\sum_{j=i+1}^{i-1}x_j-y_i+y_{i+1},
\end{eqn}
where $i=1,\dots,n$. Thus there are $3n$ such missing letters for an even-$n$-gon. A more substantial example is the box diagram, whose missing letters are listed in table \ref{table:evengon_letters}.

These examples should be viewed as evidence rather than as a final classification. Nevertheless, they already point to a coherent picture: integration does not erase the simplicity seen in the time-domain representation. Instead, it translates it into restrictions on the alphabet, with missing letters organized by graph-theoretic data such as leaves, gray vertices, and partial tubings. The natural next step is to turn these examples into an intrinsic rule for the correlator alphabet, a point to which we return in the conclusion.

\section{Conclusions and Outlook}
In this paper we developed a direct approach to conformally coupled $\phi^3$ correlators in de Sitter space. Starting from the momentum-space dressing rules, we inverted the construction and obtained a time integral representation whose elementary objects are flat space correlators with both field and conjugate momentum insertions. This mixed basis is the central structural point of the paper. It is not merely a convenient rewriting of the wavefunction story; rather, it is the basis in which the correlator closes under the graph operations that naturally arise before the final energy integrations are performed.

Several simplifications become transparent in this representation. Graphs with an odd number of conjugate momentum insertions vanish by parity, explaining the weight drop of odd-point correlators. Tree graphs obey leaf recursions, more general topologies admit fusion recursions, and melonic subgraphs collapse to effective lower-complexity graphs. The same time-space language also makes the leading behavior near total- and partial-energy singularities manifest. In particular, the leading term is universal and reproduces the expected amplitude or factorized lower-point data, while the first subleading term vanishes. These facts suggest that at least part of the simplicity of de Sitter correlators is intrinsic to the correlator itself, and not merely inherited from the wavefunction after non-trivial cancellations.

We also took a first step beyond the integrand by studying integrated symbols and alphabets. In the tree level examples considered here, including chains and stars, the correlator alphabet is smaller than the corresponding wavefunction alphabet. The missing letters are not random: they admit a natural interpretation in terms of leaves, gray vertices, and partial tubings. At loop level the pattern becomes richer, as already illustrated by polygon examples, but the same lesson persists: the full in-in observable appears to organize its analytic structure in a language better adapted to correlators than to wavefunction coefficients.

There are several natural directions in which this picture should be sharpened. The first is to place the direct time-space representation on a more systematic foundation. Here it was obtained by inverting the dressing rules \cite{Chowdhury:2025ohm}, but it would be desirable to derive it directly from the in-out formalism \cite{Donath:2024utn}. Such a derivation should clarify the origin of the mixed field/conjugate momentum basis, the role of the dotted sector, and the precise range of validity of the construction. It would also make clear whether the basis found here is special to CC $\phi^3$ theory or part of a more general correlator-first organization, potentially extending to other interactions, masses \cite{Raman:2025tsg, Baumann:2026atn}, spinning external states \cite{Chowdhury:2025nnk}, and shadow-transform formulations \cite{Sleight:2020obc,DiPietro:2021sjt,Heckelbacher:2022hbq, Chowdhury:2023arc}.

A second direction is to turn the alphabetic evidence into a general rule. The examples in
this paper strongly suggest that the correlator has its own graph-theoretic alphabet, related
to but smaller than the wavefunction alphabet. It would be very interesting to determine
which tubings survive for an arbitrary graph, which letters are removed by the correlator
projection, and what geometric object controls this reduction after integration. A related
global question is how the graph-by-graph structure found here is reorganized after summing
over all graphs contributing to a fixed observable. Our analysis suggests that individual
graph correlators already know about a reduced alphabet, with missing letters controlled by
leaves, gray vertices and partial tubings. It would be valuable to understand whether these
graph-wise reductions assemble into a positive-geometric description of the full correlator,
along the lines of cosmohedra, graph correlahedra, the correlatron, amplitubes and in-in
correlator geometries
\cite{Arkani-Hamed:2024jbp,Figueiredo:2025daa,Ardila-Mantilla:2026cbo}. Closely related is the question of differential equations.
Recent work on kinematic flow and canonical differential equations suggests that cosmological
observables may be governed by rigid systems of differential equations
\cite{Arkani-Hamed:2023decorr,Arkani-Hamed:2023kflow,Hang:2024xas,
Baumann:2024loopflow,Baumann:2025qjx,Capuano:2025cde,Baumann:2026atn}.
The enlarged black/white system introduced here may provide a natural correlator basis for
such equations, and could give a direct route to symbols, adjacency constraints, and possible
cluster-algebraic structures
\cite{Mazloumi:2025pmx,Capuano:2025myy,Paranjape:2026htn,Capuano:2026pgq}.

A third direction concerns loops, renormalization, and unitarity. Even the simple bubble example shows that the integrated loop answer mixes the formal simplicity visible at the integrand level with the familiar subtleties of regularization. A systematic treatment should identify regulators and subtraction schemes that preserve as much of the correlator-first structure as possible, disentangle physical singularities from regulator artifacts, and clarify how ultraviolet effects are reorganized in this language \cite{Lee:2023jby,Chowdhury:2026upp,Nowinski:2025cvw,Farren:2026hao}. It would also be important to connect the direct representation more explicitly to cuts, discontinuities, and dispersion-like formulae \cite{Melville:2021lst,Chowdhury:2026upp}. This may ultimately lead to a loop-level cosmological bootstrap formulated directly for renormalized correlators, with amplitudes, cuts, and energy singularities entering on equal footing.

Finally, it would be worthwhile to test how far the correlator-first viewpoint extends beyond the present model. The most immediate generalizations are to other scalar interactions, spinning correlators, and broader classes of cosmological observables. More conceptually, the results here raise the possibility that the equal-time correlator is not just the final observable obtained after processing the wavefunction, but a primary object with its own geometry, alphabet, differential equations, and singular limits. Establishing the scope of this statement, or finding its precise limitations, would sharpen our understanding of the analytic structure of quantum fields in cosmological spacetimes.

\subsection*{Acknowledgements}
CC thanks Joe Marshall for many discussions on this subject and ongoing collaborations and Ross Glew for clarifying some aspects of the energy expansions. We also thank Qu Cao and Yuyu Mo for useful discussions. C.C. is supported by an
STFC consolidated grant (ST/X000583/1). S.H. has been supported by the National Natural Science Foundation of China under Grant No. 12225510, 12447101, and by the New Cornerstone Science Foundation.

\begin{appendix}
\section{Loop Integration}\label{app:loop}
We use the dressing rules of section \ref{sec:simplicity} to evaluate the loop integral for the bubble diagram for generic momentum. Loop corrections have been extensively studied in dS including \cite{Chowdhury:2023khl, Cohen:2024anu, Chakraborty:2025mhh, Melville:2021lst,  Jain:2025maa, Nowinski:2025cvw, Farren:2026hao, Lee:2023jby, Salcedo:2022aal, Creminelli:2024cge, Benincasa:2024ptf, Pimentel:2026kqc}. For the purpose of this appendix we are only concerned with the nature of the loop integrals and postpone a systematic discussion of their renormalization to future work. We regularize the loop momentum with a hard cutoff on the comoving momentum \cite{Weinberg:2005vy}. In general one must exercise caution when regularizing loop integrals, as this can lead to violations of conformal invariance \cite{Senatore:2009cf, Chowdhury:2023arc}. 

Consider the bubble graph formed by the dashed propagators in $\phi^3$ theory, 
\begin{eqn}
\begin{tikzpicture}[baseline]
\draw (-1.5, 0) -- (-0.5, 0);
\draw (1.5, 0) -- (0.5, 0);
\draw (0,0) circle (0.5);
\draw[dashed] (-0.5, 0) -- (0, 1);
\draw[dashed] (0.5, 0) -- (0, 1);
\end{tikzpicture} &= \intinf \frac{\mathrm{d}p}{(p^2 + k^2)^2} \int \frac{\mathrm{d}^4 L}{L^2(L + K)^2}
\end{eqn} 
Upon evaluating the flat space loop integral with a cutoff $\Lambda$ we obtain,
\begin{eqn}
\begin{tikzpicture}[baseline]
\draw (-1.5, 0) -- (-0.5, 0);
\draw (1.5, 0) -- (0.5, 0);
\draw (0,0) circle (0.5);
\draw[dashed] (-0.5, 0) -- (0, 1);
\draw[dashed] (0.5, 0) -- (0, 1);
\end{tikzpicture} &= \intinf \mathrm{d}p \log^2\left(\frac{p + \mathrm{i} k}{p - \mathrm{i} k} \right) \log \left( \frac{p^2 + k^2}{\Lambda^2} \right) \\
&= -\frac{4}{3} \pi  k \left[12 \log 2 \log \left(\frac{k}{\Lambda }\right)+\pi ^2+6 (\log 2-2) \log2\right]
\end{eqn}
The divergence can be renormalized by using the standard counterterm as the one in flat space \cite{Heckelbacher:2022hbq, Chowdhury:2023arc}.
Similarly for the dotted diagram we get
\begin{eqn}
\begin{tikzpicture}[baseline]
\draw (-1.5, 0) -- (-0.5, 0);
\draw (1.5, 0) -- (0.5, 0);
\draw (0,0) circle (0.5);
\draw[dotted] (-0.5, 0) -- (0, 1);
\draw[dotted] (0.5, 0) -- (0, 1);
\end{tikzpicture} = \intinf \mathrm{d}p \log\left( \frac{p^2 + k^2}{\Lambda^2} \right) 
\end{eqn}
which results in a divergent integral. This implies that additional counterterms are required to renormalize these divergences and we hope to comment on these in a future work. 

For generic external energies we instead have,
\begin{eqn}
\int_{X_1}^{\infty} dx_1 \int_{X_2}^{\infty} dx_2 \begin{tikzpicture}[baseline]
\draw (0,0) circle (0.5);
\node at (-0.5, 0) {$\bullet$};
\node at (0.5, 0) {$\bullet$};
\draw[dashed] (-0.5, 0) -- (0, 1);
\draw[dashed] (0.5, 0) -- (0, 1);
\node at (-0.65, 0) {$x_1$};
\node at (0.65, 0) {$x_2$};
\end{tikzpicture} &= \intinf \mathrm{d}p \log\left(\frac{p + \mathrm{i} x_1}{p - \mathrm{i} x_1} \right) \log\left(\frac{p + \mathrm{i} x_2}{p - \mathrm{i} x_2} \right) \log \left( \frac{p^2 + k^2}{\Lambda^2} \right) 
\end{eqn}
The integral results in a lengthy expression with Polylogs of weight 2.

\end{appendix}

\newpage
\bibliographystyle{JHEP}
\bibliography{references}
\end{document}